\documentclass[conference]{IEEEtran}
\usepackage{geometry}
\usepackage{graphicx}%
\usepackage{subfigure}
\usepackage{setspace}
\usepackage{amsmath}
\usepackage{color}

\begin{document}
%\IEEEoverridecommandlockouts

% \title{\Large\textbf A Noise Filter for Dynamic Vision Sensors based on Global Space and Time information}%\\~\\
% 	% \large\textbf Preparation of Papers in Two-Column Format\\
% 	% for the ASP-DAC 2020 (\LaTeX version)}
%
% 	\author{\normalsize
%   \begin{tabular}[t]{c@{\extracolsep{2em}}c@{\extracolsep{2em}}c@{\extracolsep{2em}}c@{\extracolsep{2em}}c}
% 	% \begin{tabular}[t]{c@{\extracolsep{8em}}c}
% 		\large Shasha Guo& \large Ziyang Kang& \large Lei Wang& \large Limeng Zhang& \large Xiaofan Chen& \large Shiming Li& \large Weixia Xu \\
% 		\\
%     \multicolumn{5}{c}{College of Computer Science and Technology} \\
%     \multicolumn{5}{c}{National University of Defense Technology} \\
%     \multicolumn{5}{c}{Changsha, Hunan, China. 410073} \\
%     \multicolumn{5}{c}{e-mail: {guoshasha13, kangziyang14}@nudt.edu.cn} \\
% 		% Author Department & Coauthor Department \\
% 		% Author Institute  & Coauthor Institute \\
% 		% City, ST~~zipcode & City, ST~~zipcode\\
% 		% Tel: 123-456-7890 & Tel: +81-3-333-1234\\
% 		% Fax: 123-456-0987 & Fax: +81-3-333-5678\\
% 		% e-mail: aaa@bbb.ccc.ddd & e-mail eee@ffff.ggg.hh\\
% 		\end{tabular}}
% 		% use for special paper notices
% 		%\IEEEspecialpapernotice{(Invited Paper)}
%
% 		% make the title area
% 		\maketitle

\title{\Large A Noise Filter for Dynamic Vision Sensors \\ using Self-adjusting Threshold}%\\~\\
	% \large\textbf Preparation of Papers in Two-Column Format\\
	% for the ASP-DAC 2020 (\LaTeX version)}

	\author{\normalsize
  \begin{tabular}[t]{c@{\extracolsep{1em}}c@{\extracolsep{1em}}c@{\extracolsep{1em}}c@{\extracolsep{1em}}c@{\extracolsep{1em}}c@{\extracolsep{1em}}c}
	% \begin{tabular}[t]{c@{\extracolsep{8em}}c}
		\large Shasha Guo& \large Ziyang Kang& \large Lei Wang& \large Limeng Zhang& \large Xiaofan Chen& \large Shiming Li& \large Weixia Xu \\
		\\
    \multicolumn{7}{c}{College of Computer Science and Technology} \\
    \multicolumn{7}{c}{National University of Defense Technology} \\
    \multicolumn{7}{c}{Changsha, Hunan, China. 410073} \\
    \multicolumn{7}{c}{e-mail: {guoshasha13, kangziyang14}@nudt.edu.cn} \\
		% Author Department & Coauthor Department \\
		% Author Institute  & Coauthor Institute \\
		% City, ST~~zipcode & City, ST~~zipcode\\
		% Tel: 123-456-7890 & Tel: +81-3-333-1234\\
		% Fax: 123-456-0987 & Fax: +81-3-333-5678\\
		% e-mail: aaa@bbb.ccc.ddd & e-mail eee@ffff.ggg.hh\\
\end{tabular}}

\maketitle

\makeatletter
\def\ps@IEEEtitlepagestyle{%
  \def\@oddfoot{\mycopyrightnotice}%
  \def\@evenfoot{}%
}
\makeatother
\def\mycopyrightnotice{%
  \begin{minipage}{\textwidth}
    \footnotesize
    \hfill\\~\\
  \end{minipage}
  \gdef\mycopyrightnotice{}% just in case
}

{\small\textbf Abstract---
 %Abstract is a brief (50-80 word) synopsis of your paper.
% The purpose is to provide a quick outline of your
% presentation, giving the reader an overview of the
% research. It must be fit within the size allowed, which is
% about 3 inches or 7.5 centimeters.
Neuromorphic event-based dynamic vision sensors (DVS) have much faster sampling rates and a higher dynamic range than frame-based imagers. However, they are sensitive to background activity (BA) events which are unwanted.
we propose a new criteria with little computation overhead for defining real events and BA events by utilizing the global space and time information rather than the local information by Gaussian convolution, which can be also used as a filter. We denote the filter as GF. We demonstrate GF on three datasets, each recorded by a different DVS with different output size. The experimental results show that our filter produces the clearest frames compared with baseline filters and run fast.}

% \footnotetext[1]{Shasha Guo and Ziyang Kang contribute to this article equally.}
% %\vspace{-0.2in}

% dynamic vision sensor \sep background noise filter \sep spatiotemporal correlation

% \maketitle

\section{Introduction}

Research on neuromorphic event-based sensors (``silicon retinae'') started a few decades back \cite{mead91}. Recently, the technology has matured to a point where there appears some commercially available sensors. Some of the popular dynamic vision sensors (DVS) are DVS128 \cite{dvs128}, the Dynamic and Active pixel Vision Sensor (DAVIS) \cite{davis}, Asynchronous Time-based Image Sensor (ATIS) \cite{atis}, and the CeleX-IV \cite{bib:CeleX}.

Different from conventional frame-based imagers that work by sampling the scene at a fixed temporal rate (typically 30 frames per second), these sensors detect dynamic changes in illumination. This results in a higher dynamic range, higher sampling rate, and lower power consumption.
These sensors have several possible applications.

However, these sensors will produce background activity (BA) events under constant illumination, which are caused by temporal noise and junction leakage currents \cite{dvs128,Bs2filter,phdthesis}.
There are already multiple noise filtering methods for event-based data available. The Nearest Neighbor (NNb) filter based on spatiotemporal correlation \cite{Bs1filter, filterIeng, Bs2filter, filter2015, filter2016, feng2020event} is the most commonly employed method. Besides, there are some variations of NNb filters as well as some other filters based on differing polarity, refractory period and inter-spike interval \cite{filter2016}.

However, it is difficult to distinguish whether an event is a real event or noise with only the event itself. When the track of the target is known, the time correlation of events generated by a single pixel can be counted through repeated recording. Higher correlation suggest higher probability of real events, and vice versa. However, in the real world, it is unlikely to obtain every target’s motion information in advance, and relying on this method of judging image quality is not feasible.
Although \cite{filter2016} and \cite{feng2020event} introduce their criterion for real events, the computation is heavy and time-consuming. \cite{filter2016} use Gaussian convolution on the time diemension to obtain the correlation. And \cite{feng2020event} strengthen it by using convolution on both space dimension and time dimension. These methods make each event participate in multiple calculations in addition to the key comparison operations for judging. This will put a lot of burden on the calculation and lead to the increase of processing time.

To tackle these challenges, we design a new criteria with little computation overhead for defining real events and BA events by utilizing the global space and time information rather than the local information by Gaussian convolution. And it is naturally that this can be used as a filter since it decides whether an event is the real event or the BA event and thus decides whether to pass the event or filter the event. Therefore, we introduce a criteria for DVS BA events filtering as well as a new spatiotemporal BA filter.

Our contributions are as follows. First, we propose a criteria for defining real events and BA events with little computation overhead, which is also a BA filter and called GF. Second, we demonstrate GF on three datasets, each recorded by a different DVS system with different output size. The experimental results show that our filter produces the clearest frames compared with baseline filters.
%\vspace{-0.25in}
\section{Background}
%\vspace{-0.15in}
\subsection{DVS128}
%\vspace{-0.1in}
The DVS128 \cite{dvs128} sensor is an event-based image sensor that generates asynchronous events when it detects the changes in log intensity. If the change of illumination exceeds an upper or lower threshold, the DVS128 will generate an "ON" event or "OFF" event respectively.
Each pixel independently and in continuous time quantizes local relative intensity changes to generate spike events. If changes in light intensity detected by a pixel since the last event exceed the upper threshold, pixel will generate an ON event and if these changes pass the lower threshold pixel will generate an OFF event. A pixel will not generate an event otherwise. By this mechanism, DVS128 only generates events if there is a change in light intensity, therefore, sensor’s output stream only includes the detected changes in sensed signal and does not carry any redundant data.

To encode all the event information for output, the DVS128 sensor uses Address Event Representation (AER) protocol \cite{aerprotocal} to create a quadruplets $e(p,x,y,ts)$ for each event. Specifically, $p$ is polarity, i.e., ON or OFF, x is the x-position of a pixel's event, y is the y-position of a pixel's event, and $ts$ is a 32-bits timestamp, i.e., the timing information of an event.

%\vspace{-0.23in}

\subsection{DAVIS}

DAVIS \cite{davis} combines the DVS with an active pixel sensor (APS) at the pixel level. It allows simultaneous output of asynchronous events and synchronous frames.
The DAVIS sensor also uses AER protocol to encode the event output.

\subsection{CeleX-IV}
\label{sec:CeleX}
\begin{figure}
	\centering
  \includegraphics[width=.8\linewidth]{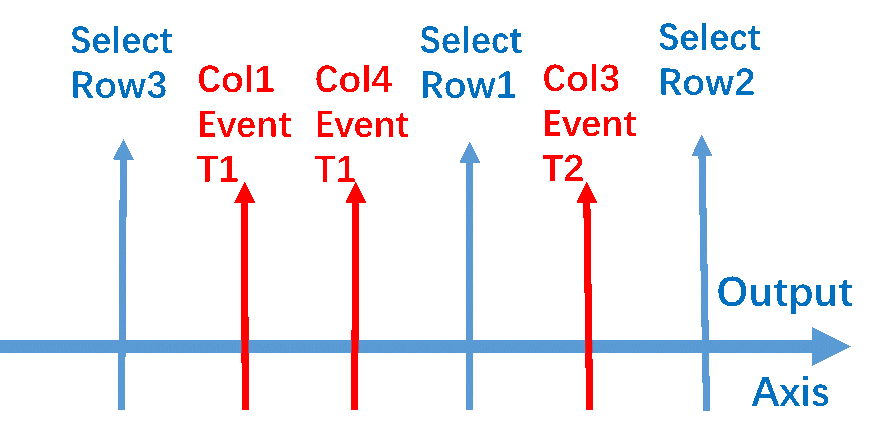}
  \caption{The event readout stream of CeleX. }
  \label{fig:CeleX}
\end{figure}

The CeleX-IV is a high resolution Dynamic Vision Sensor from CelePixel Technology Co., Ltd. \cite{bib:CeleX}. The resolution of the sensor is 768 $\times$ 640, and the maximum output rate is 200Meps. Unlike DAVIS, the event output by this sensor does not contain polarity information, but contains light intensity information of the event \cite{bib:CeleX2}, and the order of event output by this sensor follows a certain pattern. Take Fig.~\ref{fig:CeleX} as an example. First, it selects a row randomly from all the rows that has events at a single point, and then output the events on the row sequentially in a certain order. After the row finishing output, it repeats the process again from selecting a row. The key point is that the events on a row share the same timestamp.

\subsection{BA Events}
%\vspace{-0.12in}
BA events are caused due to thermal noise and junction leakage currents \cite{dvs128,Bs2filter,phdthesis}. These events degrade the quality of the data and further incurs unnecessary communication bandwidth and computing resources.
The BA and the real activity events differ in that the BA event lacks temporal correlation with events in its spatial neighborhood while the real events, arising from moving objects or changes in illumination, have a temporal correlation with events from their spatial neighbors. On the basis of this difference, the BA events can be filtered out by detecting events which do not have spatial correlation with events generated by the neighborhood pixels. Such a filter is a spatiotemporal correlation filter.
The filter decides whether an event is real or noise by checking the condition $T_{NNb} - T_{e} < dT$. If the condition is meet, the event is regarded as a real activity event.
The $T_{NNb}$ is the timestamps from the neighborhood pixels, which meet this condition: $|x_{p} - x| \leq 1$ and $|y_{p} - y| \leq 1$ where $p$ stands for a pixel. And $dT$ is the limitation for timestamp difference.
%\vspace{-0.3in}
%\vspace{-0.2in}
\section{Related work}
%\vspace{-0.12in}
\label{sec:relatedwork}
In this section, we introduce three event-based spatiotemporal filters and one frame-based filter.
There are also some other filtering methods. Researchers \cite{TNfilter} proposed a filter with neuromorphic integrate-and-fire neurons which integrate spikes not only from the corresponding pixel but also its neighborhood pixels for firing. And \cite{lifetimeestimation} assigns a lifetime to each event and the lifetime of a noise event will be assigned 0.
%\vspace{-0.22in}
% \subsection{Event-based filters}
%\vspace{-0.1in}
\label{sec:bs2}
Here we introduce three event-based filters with O(N$^2$), O(N/s) and O(N) space complexity respectively, and they will be denoted as Bs1, Bs2, and Bs3 in the rest content.

\begin{figure}
  \centering
  \subfigure[Bs1]{
  \label{fig:bs1} %% label for first subfigure
  \includegraphics[width=1.01in, height=0.8in]{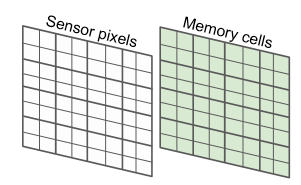}}
  \subfigure[Bs2]{
  \label{fig:bs2} %% label for second subfigure
  \includegraphics[width=1.01in, height=0.75in]{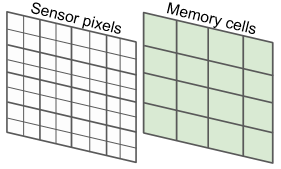}}
  \subfigure[Bs3]{
  \label{fig:bs3} %% label for second subfigure
  \includegraphics[width=1.01in, height=0.8in]{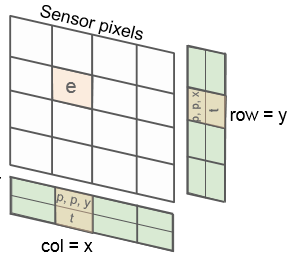}}
  %\vspace{-0.2in}
  \caption{Three event-based filters \cite{Bs3filter}.}
  \label{fig:filters}
\end{figure}

In Bs1 filter \cite{Bs1filter}, each pixel has a memory cell for storing the last event’s timestamp. The stored timestamps are used for computing the spatiotemporal correlation (Fig.~\ref{fig:bs1}).

Bs2 filter uses sub-sampling groups to reduce the memory size \cite{Bs2filter}. Each sub-sampling group of factor $s$ includes $s^{2}$ pixels and uses one memory cell for storing the timestamp of the most recent event of the group (Fig.~\ref{fig:bs2}).

Bs3 filter assigns two memory cells to each row and each column to store the most recent event in that row or column (Fig.~\ref{fig:bs3}) \cite{Bs3filter}. This filter is designed to store all the information of an event, so both the two cells are 32-bits with one for storing the timestamp and one for polarity and the other axis position.

%\vspace{-0.22in}
%\vspace{-0.25in}
\section{Proposed Filter}
%\vspace{-0.1in}
We propose a new method for separating real events and BA events by using the time difference between two events in the same pixel. By separating real events and BA events, it can naturally be used as a BA filter. It utilizes both the space information and time information from a global perspective, and we donote the filter as $GF$ for simplicity.

Table~\ref{tab:timeparameter} gives the denotations and explainations of parameters that appear in the following description.
% \begin{equation}
%   Tthr = \frac{TD \times (X + Y) \times X \times Y}{FEN \times 0.2}
%     \label{equ:CeleXtime}
% \end{equation}
% \begin{table}
%   \centering
%   \caption{Add caption}
%     \begin{tabular}{ll}
%       \hline
%     Denotation & Description \\
%     \hline
%     TD    & the time difference between the first event and the last event within a frame \\
%     FN    & the number of events of a frame \\
%     SF    & the estimated percentage of BA events in the total events of the frame \\
%     X     & the image width of the frame \\
%     Y     & the image height of the frame \\
%     s     & the subsampling window similar to Bs2 described in Section.~\textbackslash{}ref{sec:bs2} \\
%     \hline
%     \end{tabular}%
%   \label{tab:timeparameter}%
% \end{table}%

\begin{table}
  \centering
  \caption{Parameters of threshold calculation for $GF$.}
    \begin{tabular}{lp{20.835em}}
      \hline
    Para. & \multicolumn{1}{l}{Description} \\
    \hline
    TD    & the time difference between the first event and the last event within a frame \\
    ATD   & the average time difference\\
    ANEP   & the average number of events per pixel\\
    ANEM  & the average number of events per memory cell\\
    FN    & the number of events of a frame \\
    SF    & the scaling factor \\
    X     & the image width of the frame \\
    Y     & the image height of the frame \\
    s     & the subsampling window similar to Bs2 described in Section.~\ref{sec:bs2} \\
    \hline
    \end{tabular}%
  \label{tab:timeparameter}%
\end{table}%

% Where the $timeDiff$ is the time difference between the first event and the last event within a frame. The $FN$ is the number of events of a frame. $X$ and $Y$ is the image dimension of the frame. $s$ is the subsampling window similar to Bs2 described in Section.~\ref{sec:bs2}. That is, $s^{2}$ pixels share the same memory cell for storing timestamp. $SF$ is the estimated percentage of BA events in the total events of the frame.

We introduce the time threshold for the GF filter as follows, which is denoted as $TGF$.

%So we introduce the $SF$. We suppose that if there are many BA events in a frame, the number of BA events dominates the average time difference for all events in the frame. And if there are little BA events in a frame, the dominating role is the real event.
For each pixel, the $ANEP$ is
\begin{equation}
  \centering
  ANEP = \frac{FN}{X \times Y}.
\end{equation}
Intuitively, the time threshold for separating real events and BA events should be the ratio of the whole time difference and the average number of events per pixel within a frame when each pixel has a memory cell itself ($s=1$) like Bs1, which is
\begin{equation}
  \centering
  TGF = \frac{TD}{ANEP} = \frac{TD}{\frac{FN}{X \times Y}}.
\end{equation}

However, when $s^{2}$ pixels share a memory cell like Bs2, the average number of events per pixel turns to the average number of events per memory cell, $ANEM$, which is
\begin{equation}
  \centering
  ANEM = ANEP \times s^{2} = \frac{s^{2} \times FN}{X \times Y}.
\end{equation}
Thus, the time threshold for GF$_{s}$ is denoted as $TGF_{s}$, and described by Eq.~\ref{equ:dvstime0}.
\begin{equation}
  TGF_{s} = \frac{TD}{ANEM} = \frac{TD \times (X \times Y)}{s^{2} \times FN}.
    \label{equ:dvstime0}
\end{equation}

This is not the end. Since the $ATD$ of two BA events is supposed to be much larger than that of two real events according to the spatiotemporal correlation, these BA events will increase the $ATD$ between any two events in the frame compared with an ideal condition that have no BA events in the frame. In other words, it will increase the $TD$ of the frame compared with the ideal condition. The $TGF_{s}$ based on Eq.~\ref{equ:dvstime0} could be larger than expected. So we introduce the scaling factor $SF$. And the $TGF_{s}$ is updated as
\begin{equation}
  TGF_{s} = \frac{TD}{ANEM} = \frac{TD \times X \times Y}{s^{2} \times FN \times SF}.
    \label{equ:dvstime}
\end{equation}

For CeleX, due to its special timestamp assignment as described in Section.~\ref{sec:CeleX}, up to X events could have the same timestamp. We suppore that these events are regarded as one event when computing the ANEM, namely,
\begin{equation}
  \centering
  ANEM = \frac{s^{2} \times FN}{X \times (X \times Y)}.
\end{equation}

With consideration of scaling factor $SF$, the time threshold $TGF_{s}$ can be described as Eq.~\ref{equ:CeleXtime}.
\begin{equation}
  TGF_{s} = \frac{TD}{ANEM \times SF} = \frac{TD \times X \times (X \times Y)}{s^{2} \times FN \times SF}.
    \label{equ:CeleXtime}
\end{equation}
And it is worth noticing that the $TGF_{s}$ for CeleX is likely to be smaller than expected. Since it is up to X events could share the same timestamp, usually it is smaller than X.

For each frame's events, the time threshold $TGF_{s}$ is calculated based on the last frame. The first frame of one event stream is initialized as a constant number. Although the threshold is calculated based on the frame information, GF should always be seen as an event-oriented filter. The steps for the GF filter are outlined as
follows. For each event:
\begin{itemize}
  \item Fetch the corresponding memory cell and get the last recorded timestamp;
  \item Check if the present timestamp is within $TGF_{s}$ of the last timestamp. If the time difference is less than $TGF_{s}$, pass the event to the output, otherwise discard it.
  \item Store the timestamp of the new event in the corresponding memory cell.
\end{itemize}

%\vspace{-0.3in}
\section{Experiment Setup}
%\vspace{-0.17in}
\subsection{Dataset}
%\vspace{-0.12in}
We use three datasets, a collected DVS dataset, DvsGesture \cite{dataset}, a DAVIS-240C dataset Roshambo \cite{bib:RoShamBo}, and our own dataset recorded from CeleX-IV.
DvsGesture comprises 11 hand gesture categories from 29 subjects under 3 illumination conditions. Roshambo is a dataset of rock, paper, scissors and background images. We use three sub-recordings of rock, paper, and scissors. And our own dataset is also a rock-paper-scissors dataset.

To make the event stream visible, it is common to generate a picture frame from the events, either of fixed time length or of a constant number of events.
We choose to use the fixed number of events. A pixel in the picture will be 255 if it has an event. % occurred in the pixel; otherwise 0.

The baseline filters are decribed in section \ref{sec:relatedwork}.
%\vspace{-0.22in}

\subsection{Software Configuration}
The fixed time threshold used for Bs1 is 0.5 ms. For Bs2 with subsampling window $s$ ($s \textgreater 1$), the fixed time threshold is $0.5 \times (s\times s)$ ms. For Bs3, the time threshold is $\frac{0.5}{X}$ ms.
The choice of $SF$ is related to the number of events in a frame and the BA frequency under different circumstances when recording the data. We find the proper $SF$ for DVS128 and DAVIS is larger than 1 while the proper $SF$ for CeleX is smaller than 1. The $SF$ for Eq.~\ref{equ:dvstime} is set to be 10 and the $SF$ for Eq.~\ref{equ:CeleXtime} is set to be 0.2 in this work.
%\vspace{-0.27in}
\section{Experiment result}
%\vspace{-0.16in}
% \subsection{Time filter result}
First, we want to show that GF$_{x}$ well separates the real events and BA events. Then we compare the runtime cost.
\subsection{Denoising Effect}
\label{sec:visual effect}
 %As BA event frequency differs under different circumstances, we choose it based on the frequency
For Roshambo dataset, we use 5k events for generating each frame.
For our dataset, since the CeleX-IV has very large output, namely 768 $\times$ 640 pixels, we choose 50k events as the number of events per frame to make the images easy to distinguish with human eyes.

\begin{figure*}
\centering
\subfigure[INIT]{
    \begin{minipage}[t]{0.16\textwidth}
        \centering
        \includegraphics[width=0.9in]{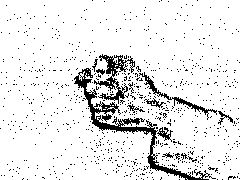}\\
        \vspace{0.02cm}
        \includegraphics[width=0.9in]{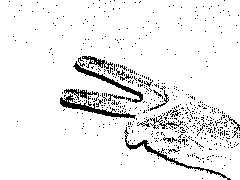}\\
        \vspace{0.02cm}
        \includegraphics[width=0.9in]{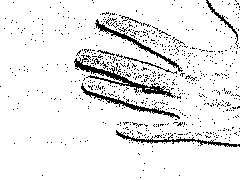}\\
        \vspace{0.02cm}
        %\caption{fig1}
    \end{minipage}%
}%
\subfigure[Bs1]{
    \begin{minipage}[t]{0.16\textwidth}
        \centering
        \includegraphics[width=0.9in]{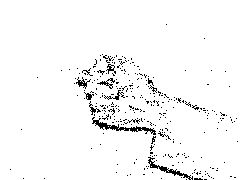}\\
        \vspace{0.02cm}
        \includegraphics[width=0.9in]{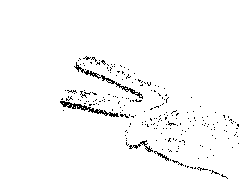}\\
        \vspace{0.02cm}
        \includegraphics[width=0.9in]{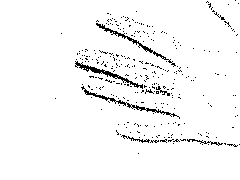}\\
        \vspace{0.02cm}
        %\caption{fig1}
    \end{minipage}%
}%
\subfigure[GF$_{1}$]{
    \begin{minipage}[t]{0.16\textwidth}
        \centering
        \includegraphics[width=0.9in]{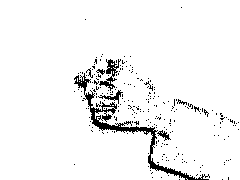}\\
        \vspace{0.02cm}
        \includegraphics[width=0.9in]{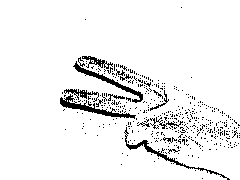}\\
        \vspace{0.02cm}
        \includegraphics[width=0.9in]{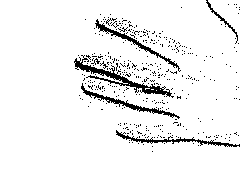}\\
        \vspace{0.02cm}
        %\caption{fig1}
    \end{minipage}%
}%
\subfigure[Bs2]{
    \begin{minipage}[t]{0.16\textwidth}
        \centering
        \includegraphics[width=0.9in]{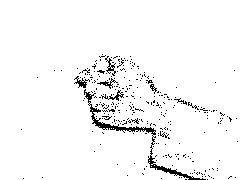}\\
        \vspace{0.02cm}
        \includegraphics[width=0.9in]{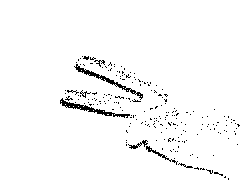}\\
        \vspace{0.02cm}
        \includegraphics[width=0.9in]{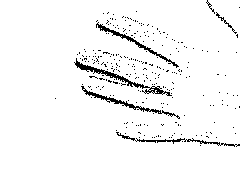}\\
        \vspace{0.02cm}
        %\caption{fig1}
    \end{minipage}%
}%
\subfigure[GF$_{2}$]{
    \begin{minipage}[t]{0.16\textwidth}
        \centering
        \includegraphics[width=0.9in]{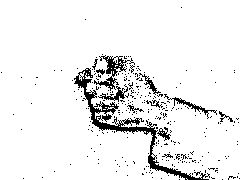}\\
        \vspace{0.02cm}
        \includegraphics[width=0.9in]{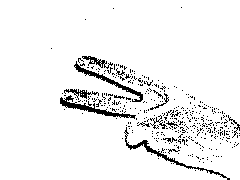}\\
        \vspace{0.02cm}
        \includegraphics[width=0.9in]{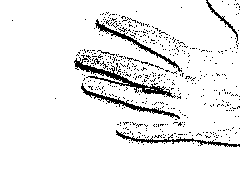}\\
        \vspace{0.02cm}
        %\caption{fig1}
    \end{minipage}%
}%
\subfigure[Bs3]{
    \begin{minipage}[t]{0.16\textwidth}
        \centering
        \includegraphics[width=0.9in]{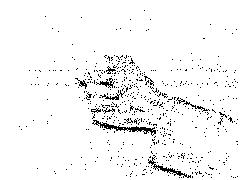}\\
        \vspace{0.02cm}
        \includegraphics[width=0.9in]{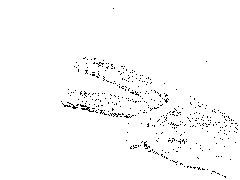}\\
        \vspace{0.02cm}
        \includegraphics[width=0.9in]{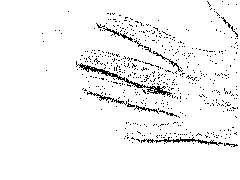}\\
        \vspace{0.02cm}
        %\caption{fig1}
    \end{minipage}%
}%
\caption{The rock-paper-scissors recorded by DAVIS240 \cite{davis}.}
\vspace{-0.2cm}
\label{fig:dvsroshambo}
\end{figure*}

Fig.~\ref{fig:dvsroshambo} shows the performance of different filters on the Roshambo dataset. It can be seen that GF$_{1}$ is clearer than Bs1 and GF$_{2}$ is clearer than Bs2. Bs3 filters the BA as well as many real events with a dim outline.

\begin{figure*}
\centering
\subfigure[INIT]{
    \begin{minipage}[t]{0.16\textwidth}
        \centering
        \includegraphics[width=0.9in]{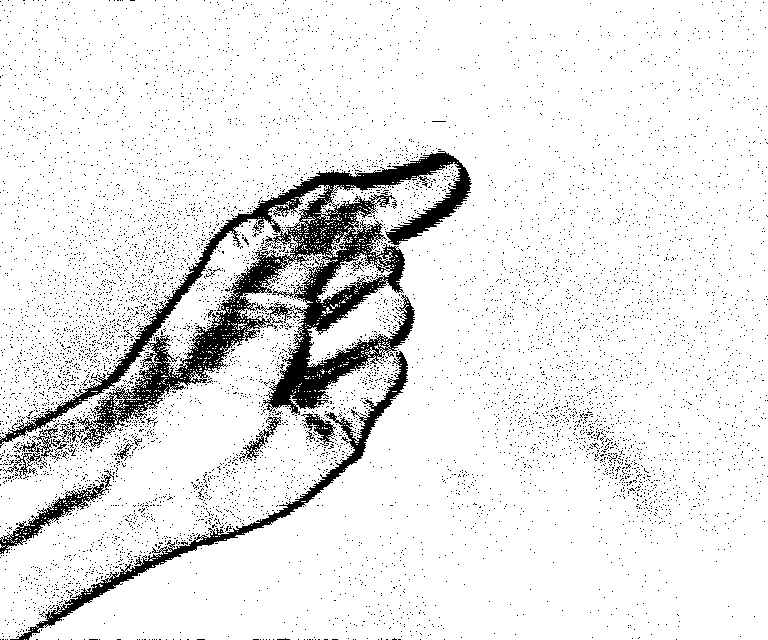}\\
        \vspace{0.02cm}
        \includegraphics[width=0.9in]{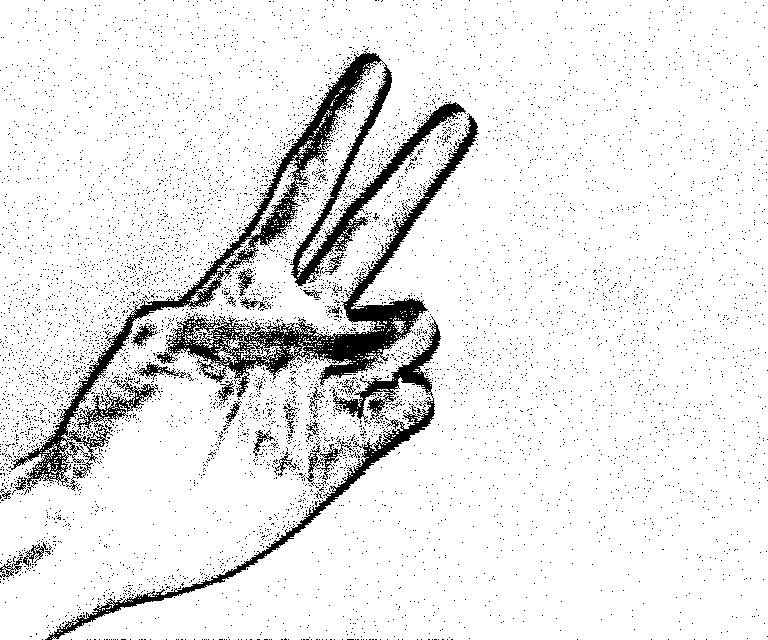}\\
        \vspace{0.02cm}
        \includegraphics[width=0.9in]{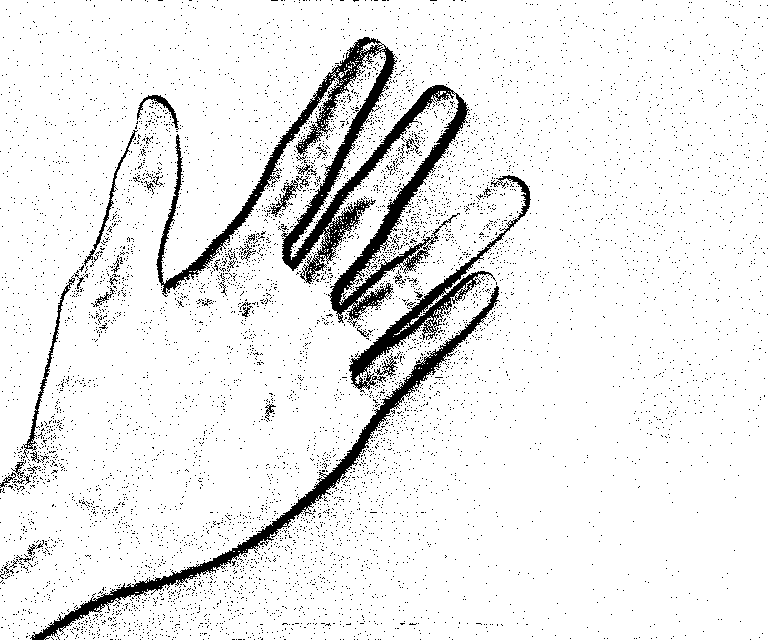}\\
        \vspace{0.02cm}
        %\caption{fig1}
    \end{minipage}%
}%
\subfigure[Bs1]{
    \begin{minipage}[t]{0.16\textwidth}
        \centering
        \includegraphics[width=0.9in]{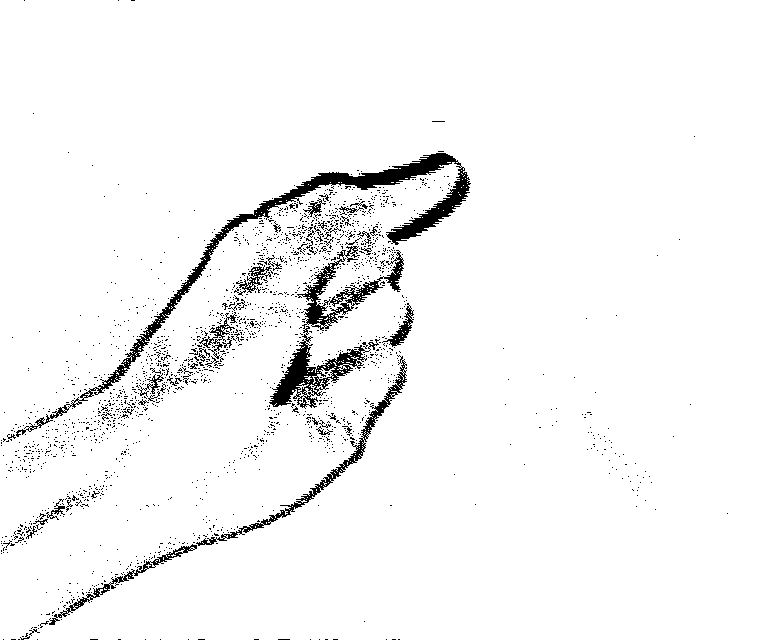}\\
        \vspace{0.02cm}
        \includegraphics[width=0.9in]{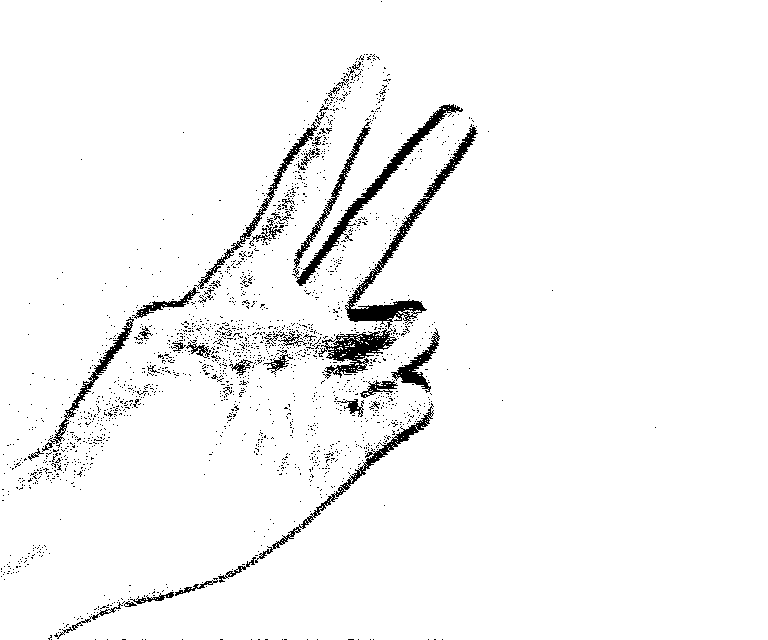}\\
        \vspace{0.02cm}
        \includegraphics[width=0.9in]{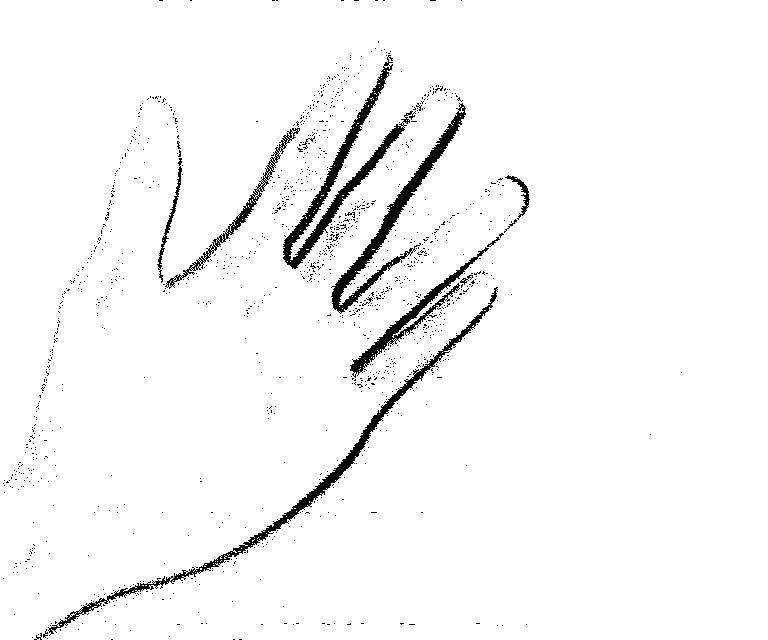}\\
        \vspace{0.02cm}
        %\caption{fig1}
    \end{minipage}%
}%
\subfigure[GF$_{1}$]{
    \begin{minipage}[t]{0.16\textwidth}
        \centering
        \includegraphics[width=0.9in]{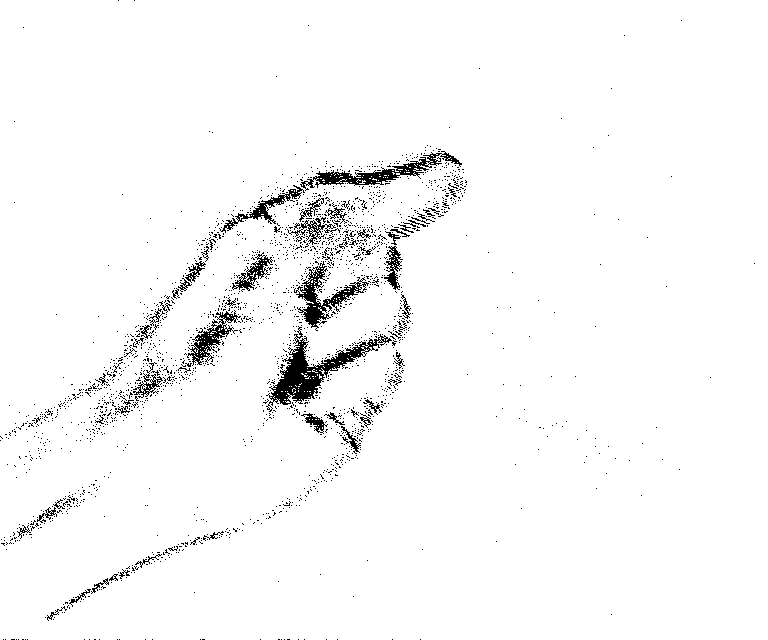}\\
        \vspace{0.02cm}
        \includegraphics[width=0.9in]{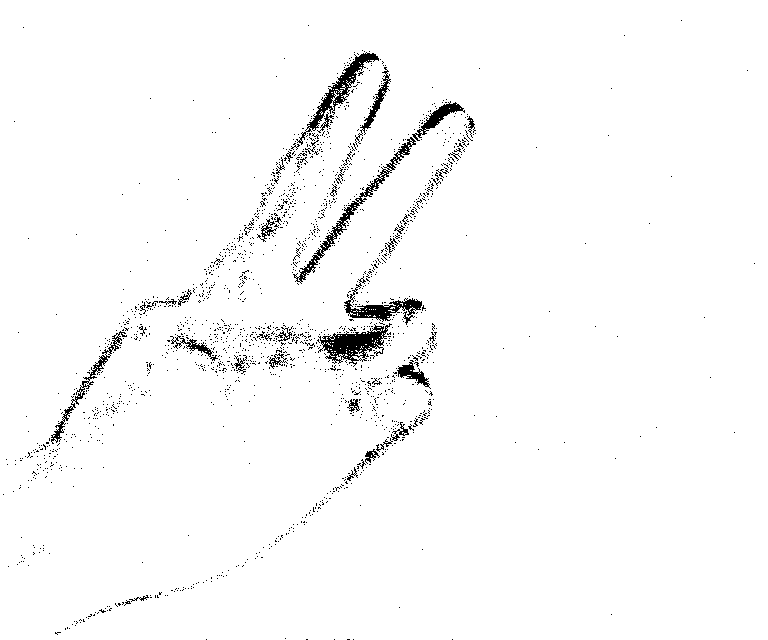}\\
        \vspace{0.02cm}
        \includegraphics[width=0.9in]{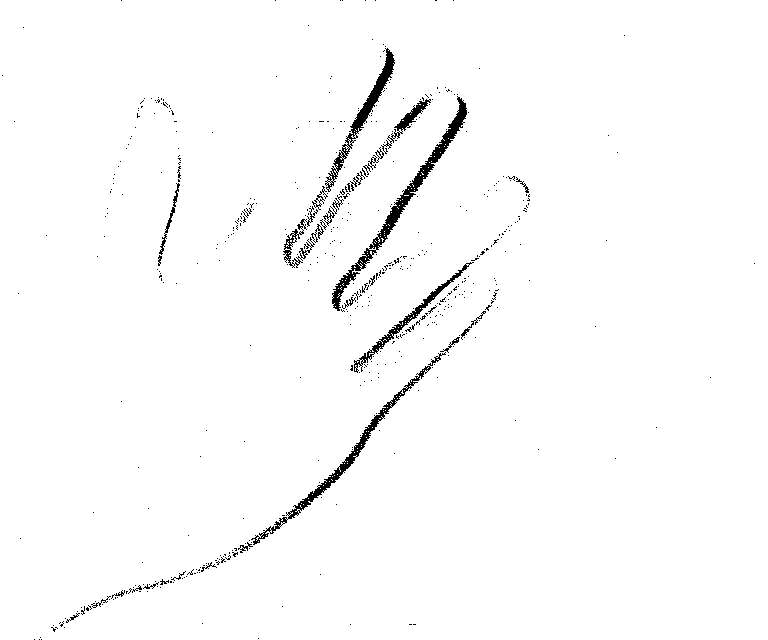}\\
        \vspace{0.02cm}
        %\caption{fig1}
    \end{minipage}%
}%
\subfigure[Bs2]{
    \begin{minipage}[t]{0.16\textwidth}
        \centering
        \includegraphics[width=0.9in]{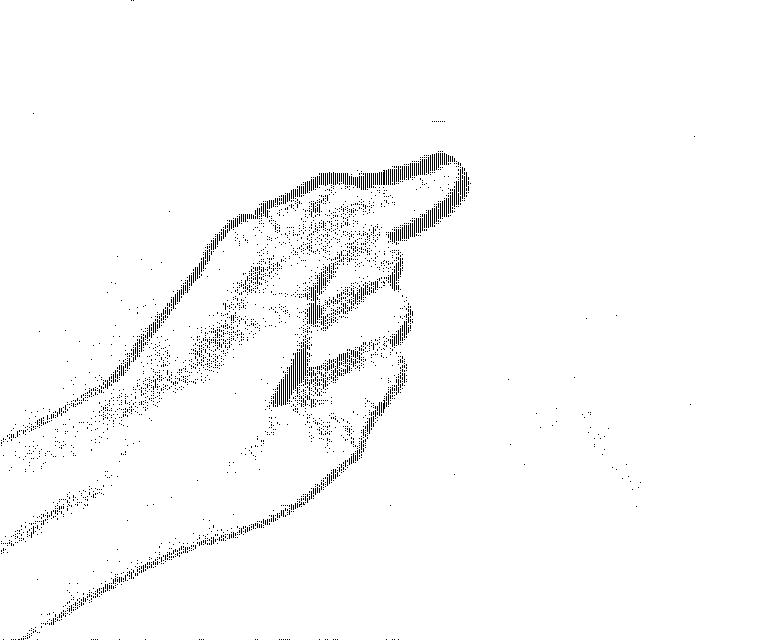}\\
        \vspace{0.02cm}
        \includegraphics[width=0.9in]{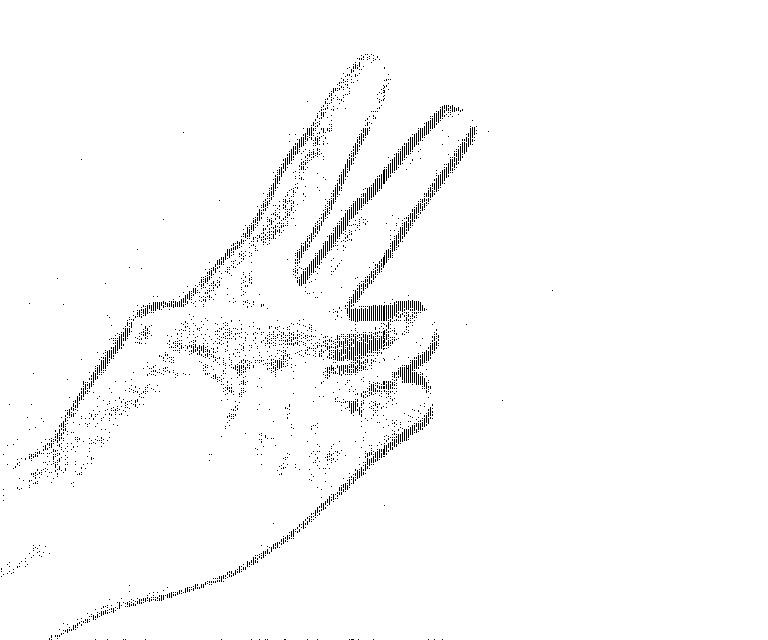}\\
        \vspace{0.02cm}
        \includegraphics[width=0.9in]{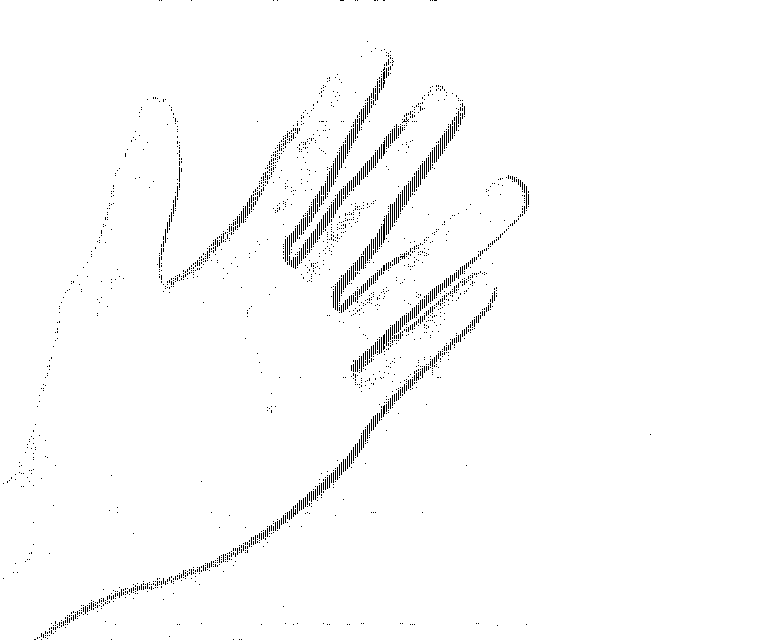}\\
        \vspace{0.02cm}
        %\caption{fig1}
    \end{minipage}%
}%
\subfigure[GF$_{2}$]{
    \begin{minipage}[t]{0.16\textwidth}
        \centering
        \includegraphics[width=0.9in]{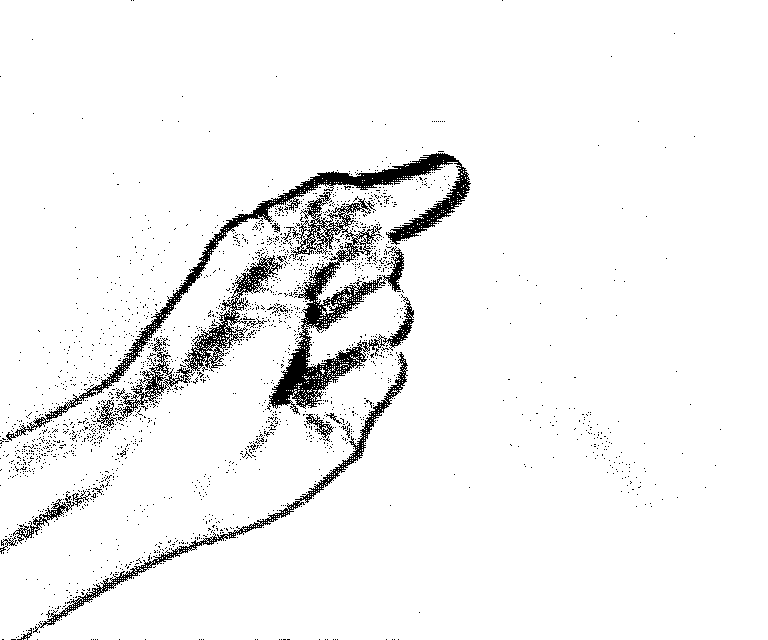}\\
        \vspace{0.02cm}
        \includegraphics[width=0.9in]{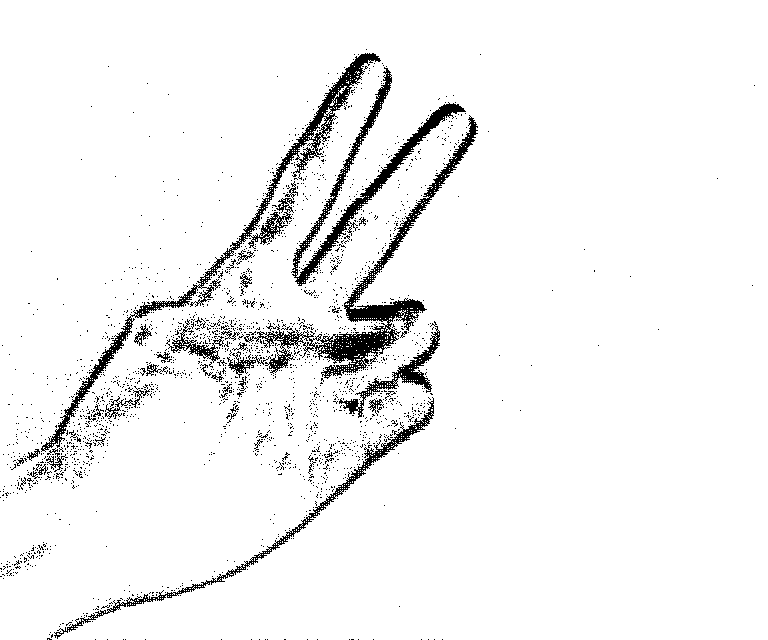}\\
        \vspace{0.02cm}
        \includegraphics[width=0.9in]{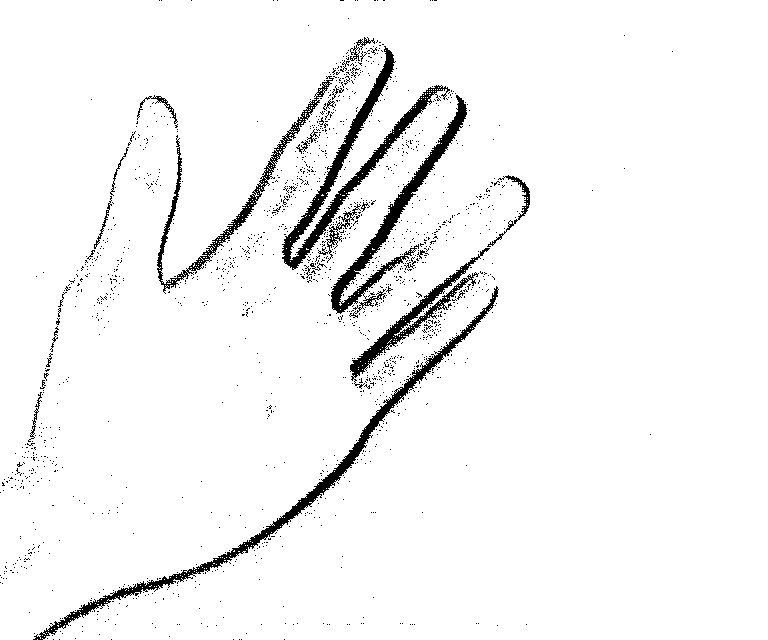}\\
        \vspace{0.02cm}
        %\caption{fig1}
    \end{minipage}%
}%
\subfigure[Bs3]{
    \begin{minipage}[t]{0.16\textwidth}
        \centering
        \includegraphics[width=0.9in]{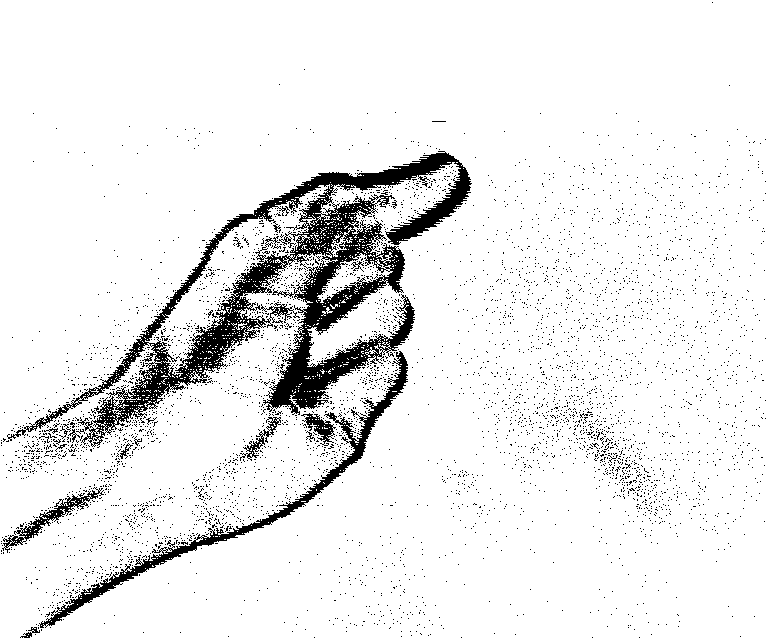}\\
        \vspace{0.02cm}
        \includegraphics[width=0.9in]{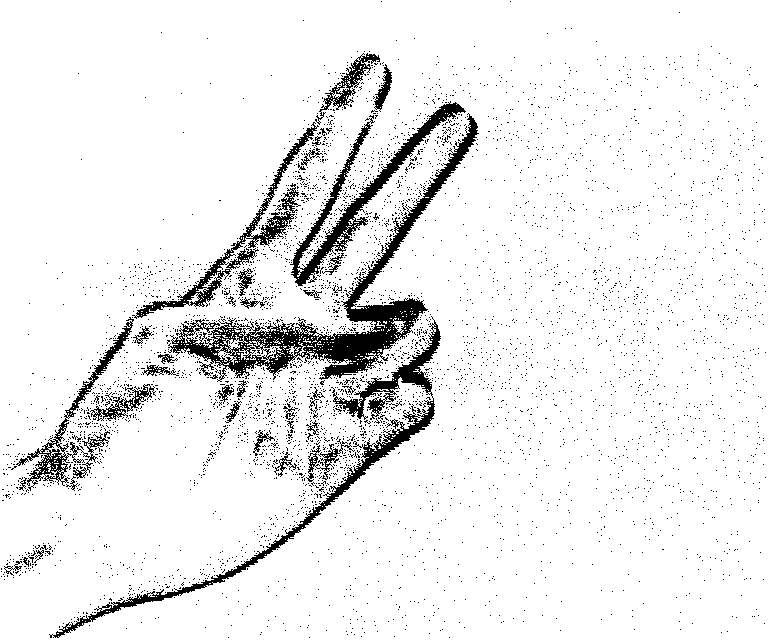}\\
        \vspace{0.02cm}
        \includegraphics[width=0.9in]{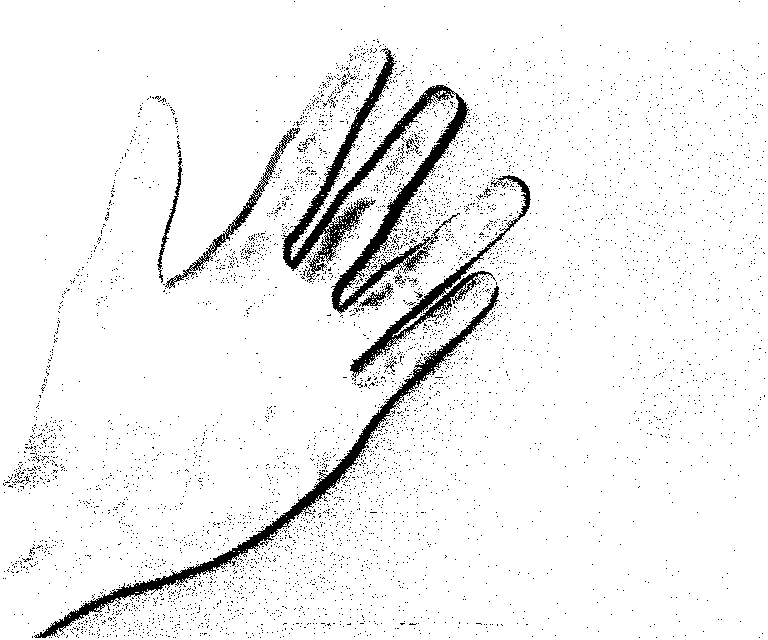}\\
        \vspace{0.02cm}
        %\caption{fig1}
    \end{minipage}%
}%
\caption{The rock-paper-scissors from CeleX when object moving fast.}
\vspace{-0.2cm}
\label{fig:roshambo1}
\end{figure*}

\begin{figure*}
\centering
\subfigure[INIT]{
    \begin{minipage}[t]{0.16\textwidth}
        \centering
        \includegraphics[width=0.9in]{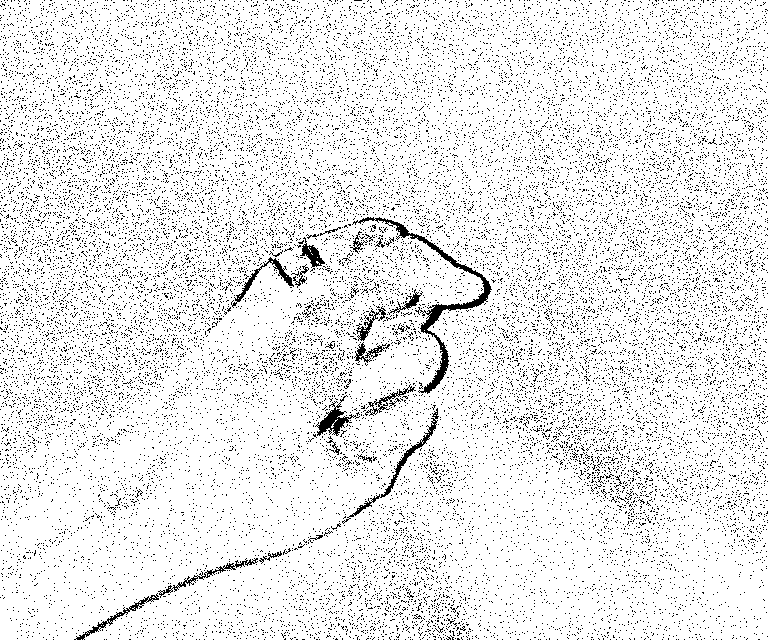}\\
        \vspace{0.02cm}
        \includegraphics[width=0.9in]{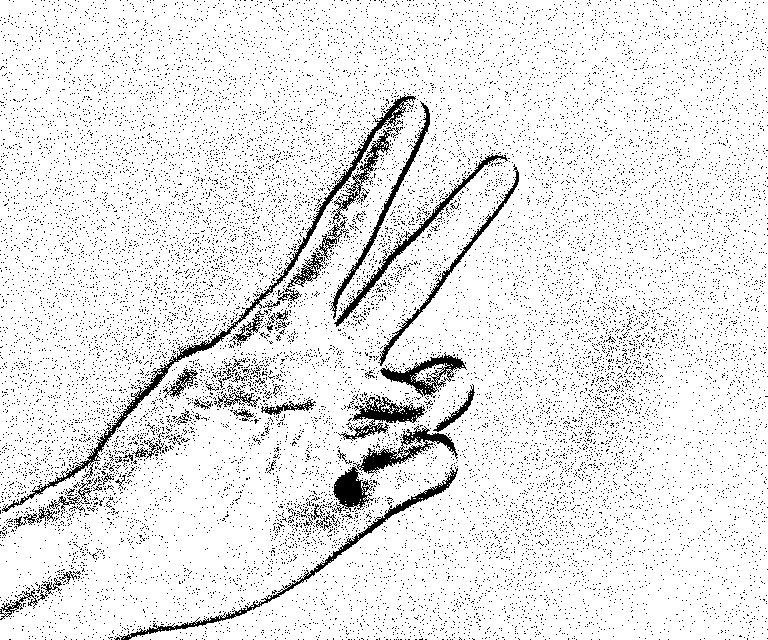}\\
        \vspace{0.02cm}
        \includegraphics[width=0.9in]{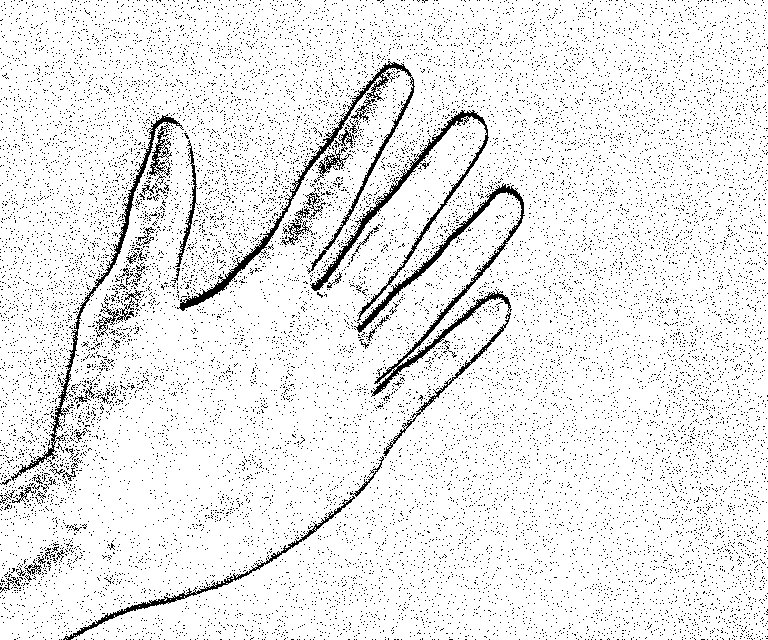}\\
        \vspace{0.02cm}
        %\caption{fig1}
    \end{minipage}%
}%
\subfigure[Bs1]{
    \begin{minipage}[t]{0.16\textwidth}
        \centering
        \includegraphics[width=0.9in]{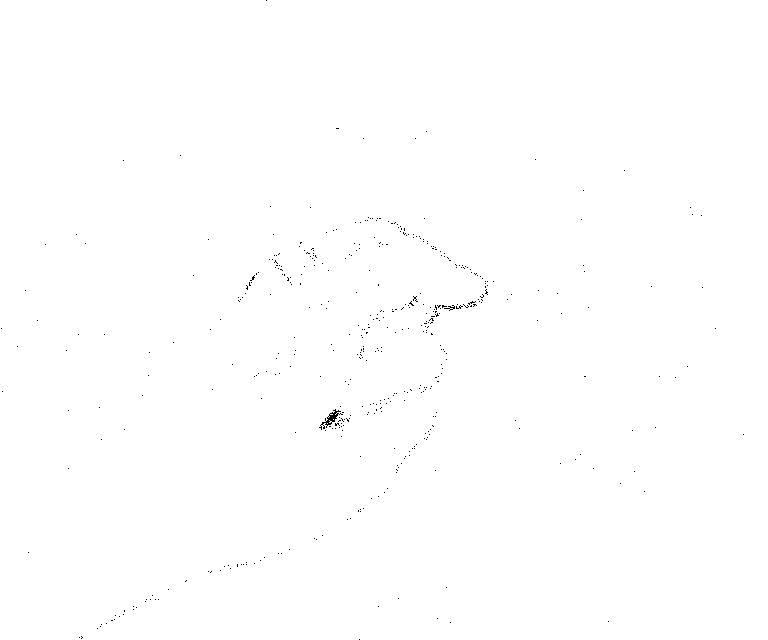}\\
        \vspace{0.02cm}
        \includegraphics[width=0.9in]{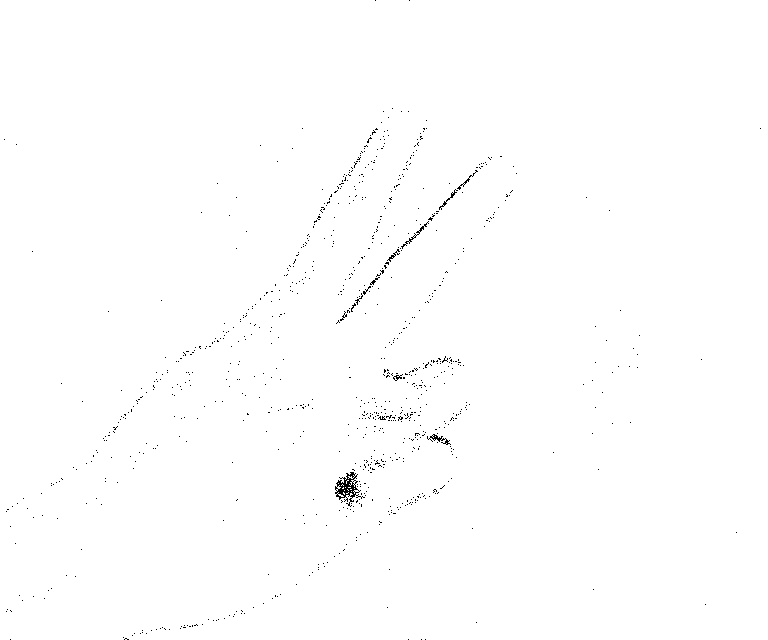}\\
        \vspace{0.02cm}
        \includegraphics[width=0.9in]{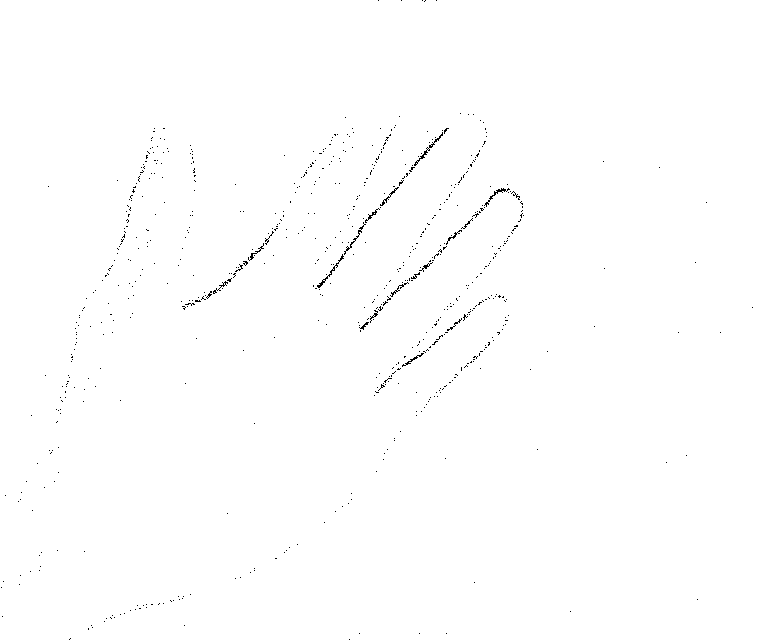}\\
        \vspace{0.02cm}
        %\caption{fig1}
    \end{minipage}%
}%
\subfigure[GF$_{1}$]{
    \begin{minipage}[t]{0.16\textwidth}
        \centering
        \includegraphics[width=0.9in]{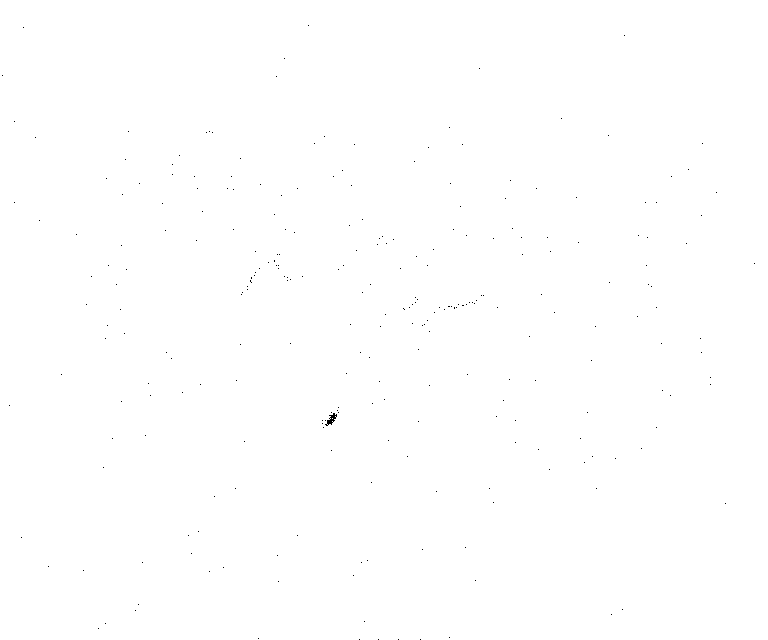}\\
        \vspace{0.02cm}
        \includegraphics[width=0.9in]{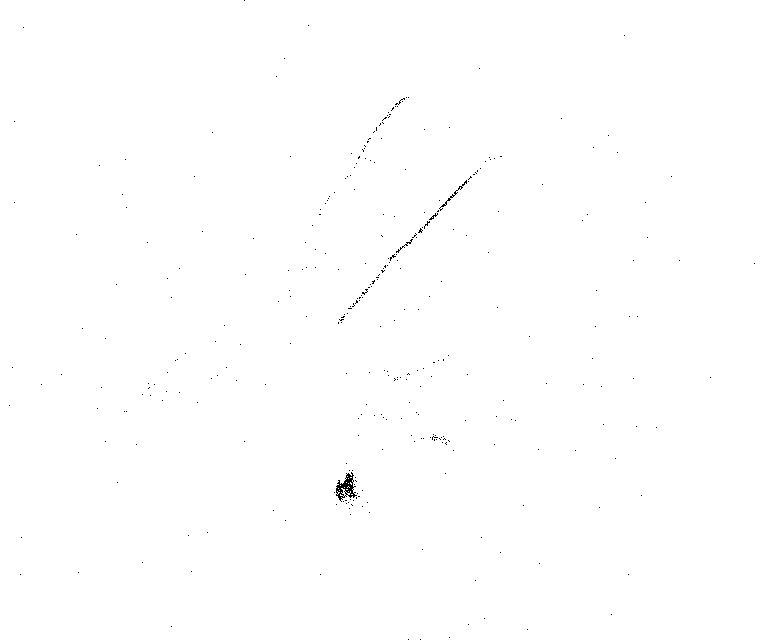}\\
        \vspace{0.02cm}
        \includegraphics[width=0.9in]{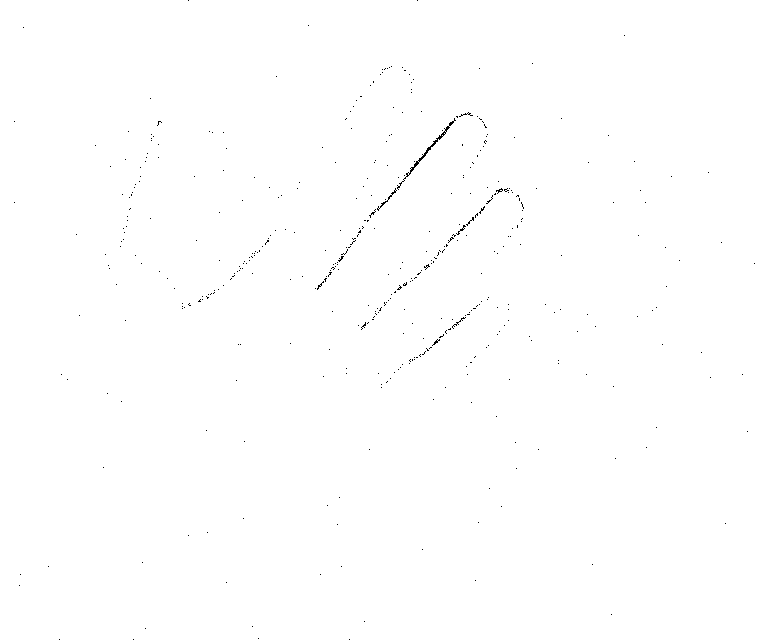}\\
        \vspace{0.02cm}
        %\caption{fig1}
    \end{minipage}%
}%
\subfigure[Bs2]{
    \begin{minipage}[t]{0.16\textwidth}
        \centering
        \includegraphics[width=0.9in]{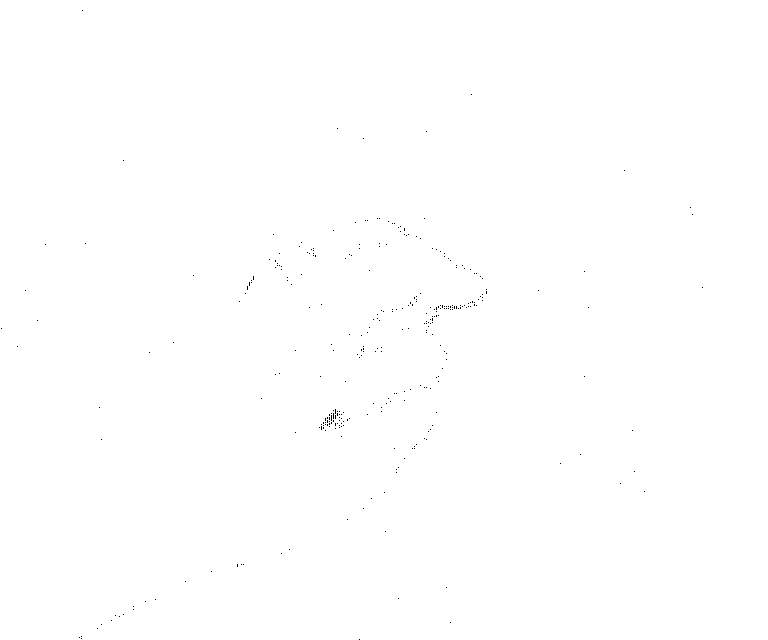}\\
        \vspace{0.02cm}
        \includegraphics[width=0.9in]{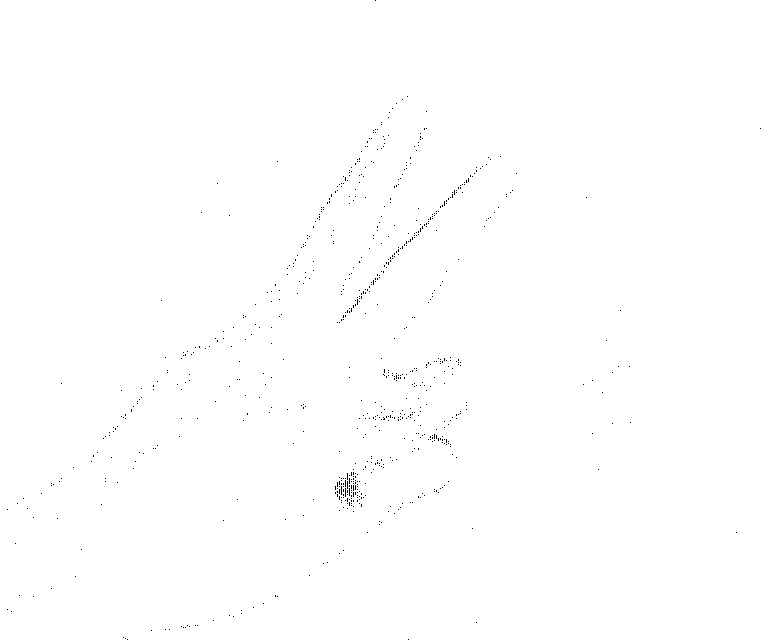}\\
        \vspace{0.02cm}
        \includegraphics[width=0.9in]{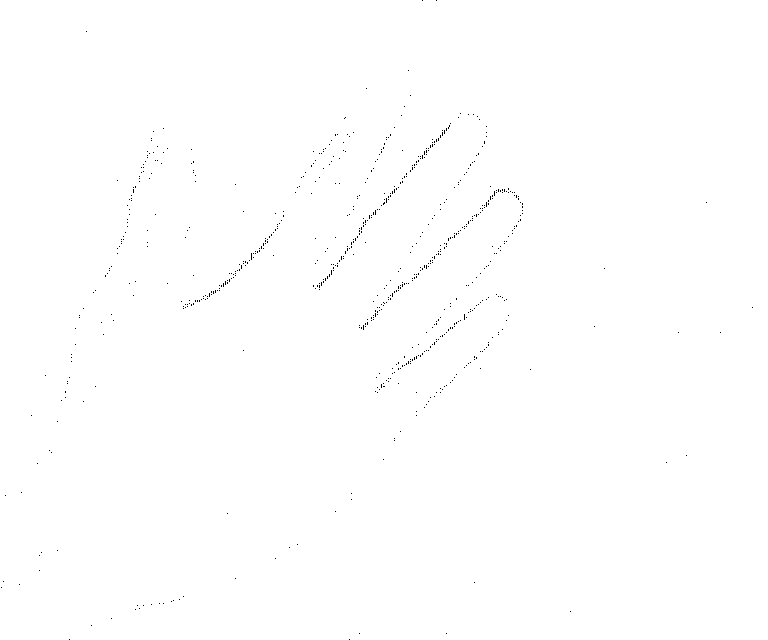}\\
        \vspace{0.02cm}
        %\caption{fig1}
    \end{minipage}%
}%
\subfigure[GF$_{2}$]{
    \begin{minipage}[t]{0.16\textwidth}
        \centering
        \includegraphics[width=0.9in]{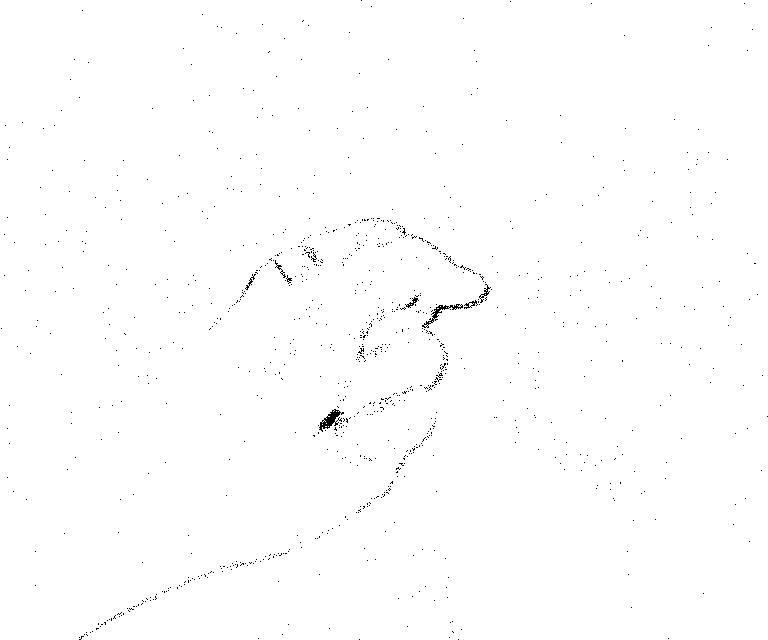}\\
        \vspace{0.02cm}
        \includegraphics[width=0.9in]{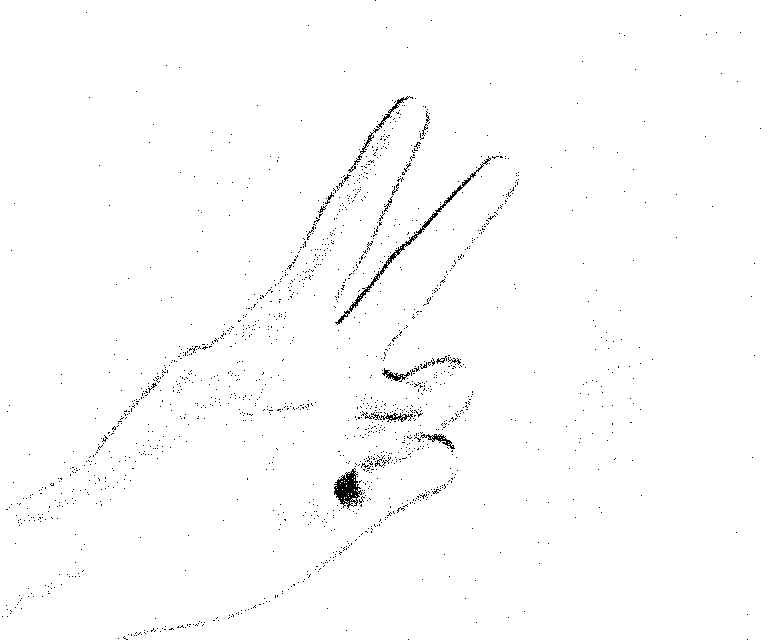}\\
        \vspace{0.02cm}
        \includegraphics[width=0.9in]{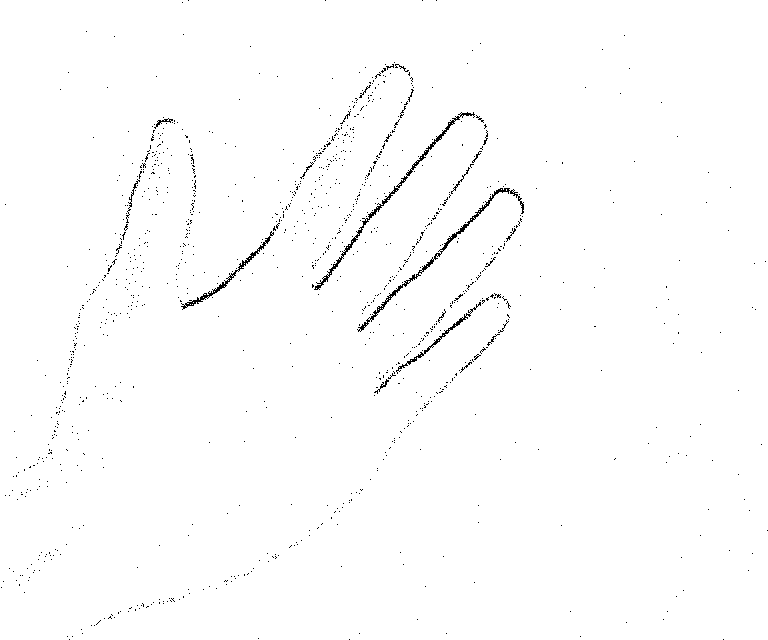}\\
        \vspace{0.02cm}
        %\caption{fig1}
    \end{minipage}%
}%
\subfigure[Bs3]{
    \begin{minipage}[t]{0.16\textwidth}
        \centering
        \includegraphics[width=0.9in]{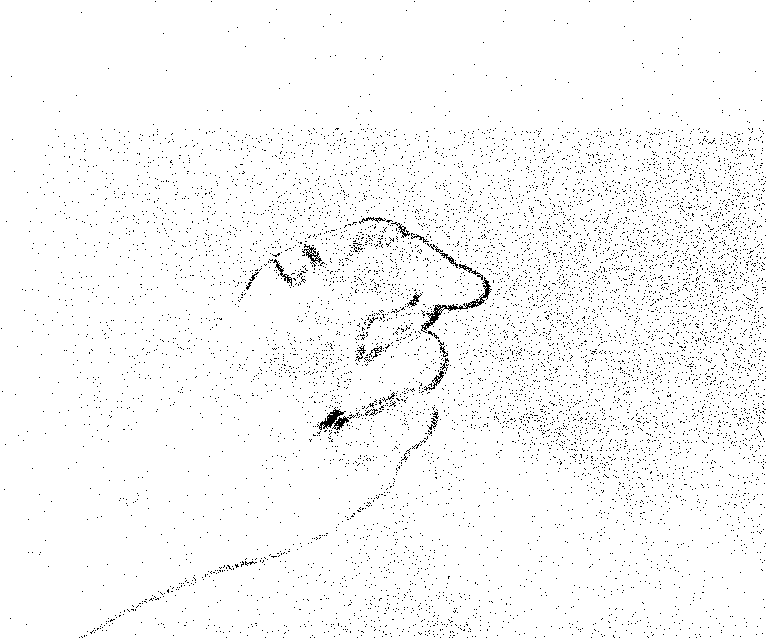}\\
        \vspace{0.02cm}
        \includegraphics[width=0.9in]{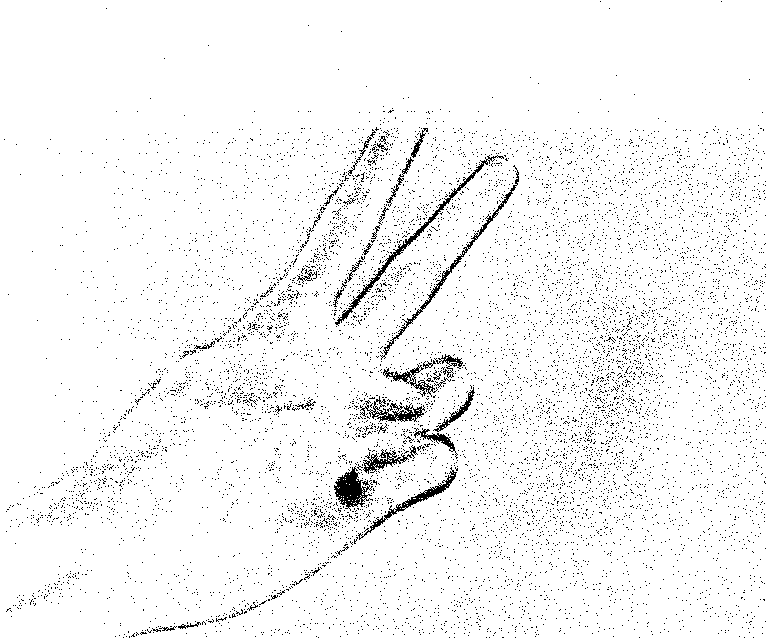}\\
        \vspace{0.02cm}
        \includegraphics[width=0.9in]{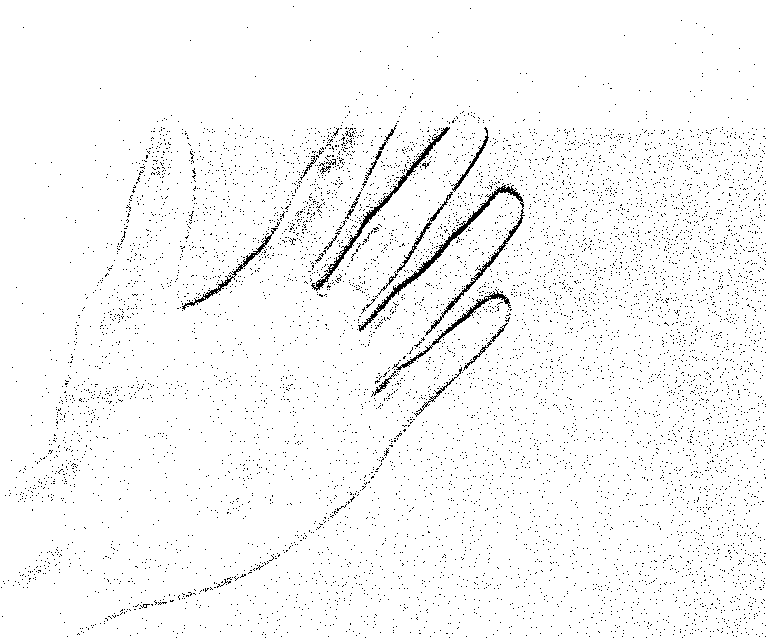}\\
        \vspace{0.02cm}
        %\caption{fig1}
    \end{minipage}%
}%
\caption{The rock-paper-scissors from CeleX when object moving slowly. The percentage of BA rises in the fixed count frames.}
\vspace{-0.2cm}
\label{fig:roshambo2}
\end{figure*}

For our dataset, we show two different cases, that is, object moving fast and moving slowly. Fig.~\ref{fig:roshambo1} depicts the events when the hand is actively moving and thus there are many real events within the frame. On the contrary, Fig.~\ref{fig:roshambo2} shows the events when the hand barely moved and thus the BA event account for a much higher percentage within the frame compared with the above cases.
It can be seen that, in Fig.~\ref{fig:roshambo1}, the initial frame don't witness many noise pixel. The effect of GF$_{1}$ is less clearer than Bs1 but clearer than Bs2. GF$_{2}$ is clearer than Bs1 and still keeps the background area clean. Bs3 is the worst as it still contains many BA noise pixels. In Fig.~\ref{fig:roshambo2}, although Bs3 shows the object, it shows many noise points as well. For the other filters, Bs1 keeps relatively more information than Bs2 and GF$_{1}$, and GF$_{1}$ and Bs2 are still very similar. GF$_{2}$ shows clear outline of the object and get rid of the noise effectively.

It can be explained why GF$_{1}$ does not keep many real events in this case. Because the object movement slows down, the time difference between real events becomes close to that between BA events. When they are close, it is hard to distinguish them using GF$_{1}$. But the background filters don't show satisfactory result as well in such cases. However, GF$_{2}$ solves this problem better because it has more space information support as a group pixels share the same memory cell while GF$_{2}$ has the same memory cost as Bs2.

\subsubsection{Quantitative Analysis based on GF}
\begin{figure}
  \centering
  \subfigure[Bs1]{
      \begin{minipage}[t]{0.24\linewidth}
          \centering
          \includegraphics[width=0.9in]{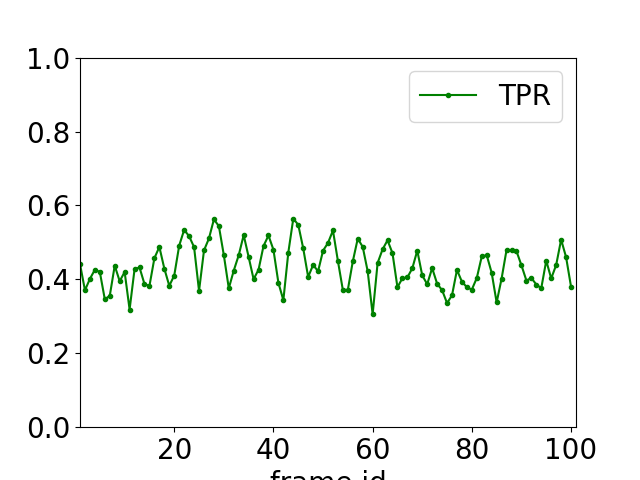}\\
          \vspace{0.02cm}
          \includegraphics[width=0.9in]{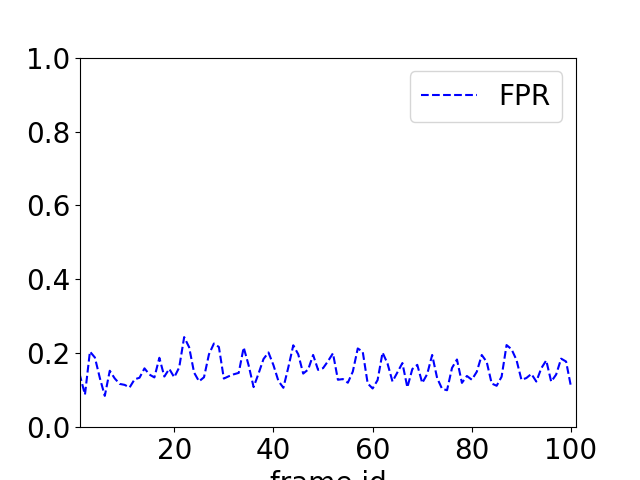}\\
          % \vspace{0.02cm}
          % \includegraphics[width=0.9in]{./Bs2_6_s2.png}\\
          \vspace{0.02cm}
          %\caption{fig1}
      \end{minipage}%
  }%
  \subfigure[Bs2]{
      \begin{minipage}[t]{0.24\linewidth}
          \centering
          \includegraphics[width=0.9in]{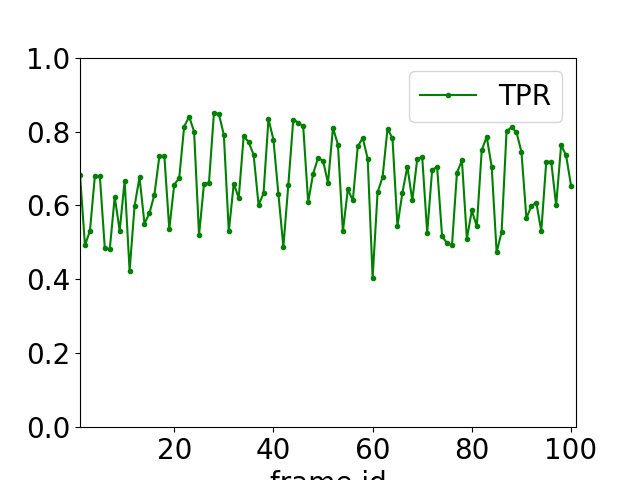}\\
          \vspace{0.02cm}
          \includegraphics[width=0.9in]{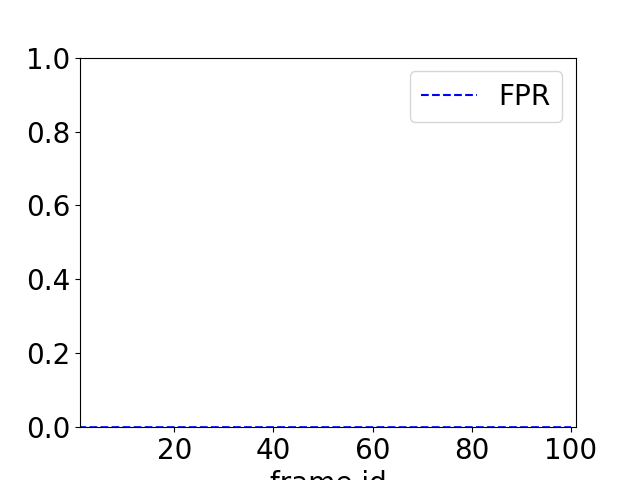}\\
          % \vspace{0.02cm}
          % \includegraphics[width=0.9in]{./TF2_6_s2.png}\\
          \vspace{0.02cm}
          %\caption{fig1}
      \end{minipage}%
  }%
  \subfigure[Bs3]{
      \begin{minipage}[t]{0.24\linewidth}
          \centering
          \includegraphics[width=0.9in]{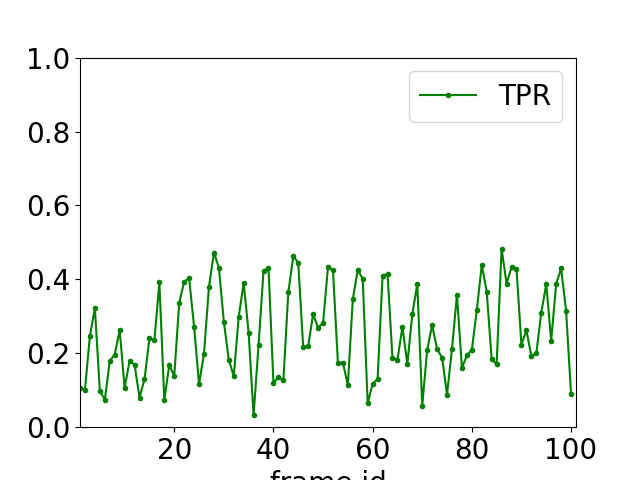}\\
          \vspace{0.02cm}
          \includegraphics[width=0.9in]{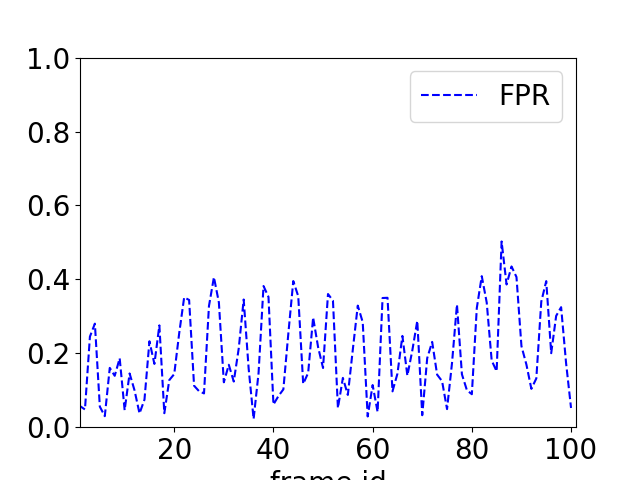}\\
          % \vspace{0.02cm}
          % \includegraphics[width=0.9in]{./Bs3_6.png}\\
          \vspace{0.02cm}
          %\caption{fig1}
      \end{minipage}%
  }%  %\vspace{-0.2in}
	\subfigure[GF1]{
      \begin{minipage}[t]{0.24\linewidth}
          \centering
          \includegraphics[width=0.9in]{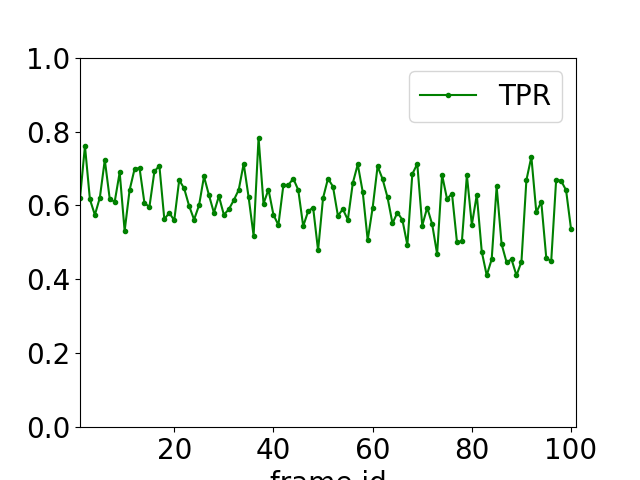}\\
          \vspace{0.02cm}
          \includegraphics[width=0.9in]{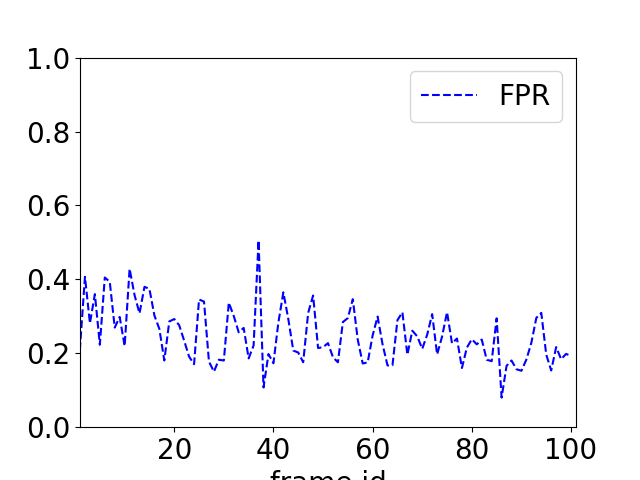}\\
          % \vspace{0.02cm}
          % \includegraphics[width=0.9in]{./Bs3_6.png}\\
          \vspace{0.02cm}
          %\caption{fig1}
      \end{minipage}%
  }%  %\vspace{-0.2in}
  \caption{Comparison of $TPR$ and $FPR$. One point in the figure represents a frame. The x-axis is the frame id. The y-axis is the ratio value ranging from 0 to 1. The top line is $TPR$, and the bottomline is $FPR$.}
  \label{fig:tfsnr}
\end{figure}

\begin{figure}
  \centering
  \subfigure[Thr=0.5ms]{
      \begin{minipage}[t]{0.24\linewidth}
          \centering
          \includegraphics[width=0.9in]{./Bs1_2.png}\\
          \vspace{0.02cm}
          \includegraphics[width=0.9in]{./Bs1_3.png}\\
          % \vspace{0.02cm}
          % \includegraphics[width=0.9in]{./Bs2_6_s2.png}\\
          \vspace{0.02cm}
          %\caption{fig1}
      \end{minipage}%
  }%
  \subfigure[Thr=1ms]{
      \begin{minipage}[t]{0.24\linewidth}
          \centering
          \includegraphics[width=0.9in]{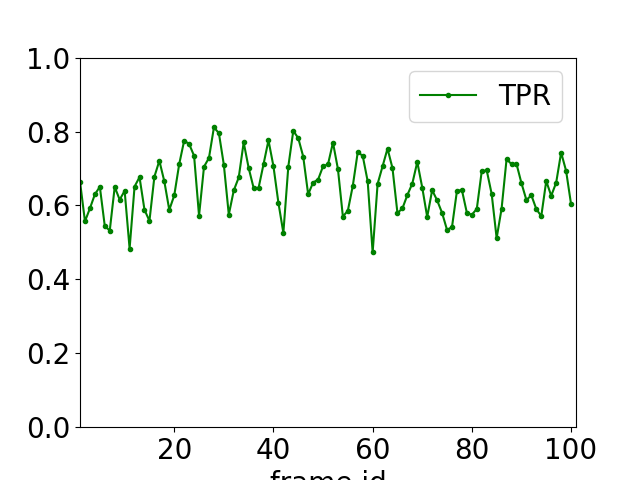}\\
          \vspace{0.02cm}
          \includegraphics[width=0.9in]{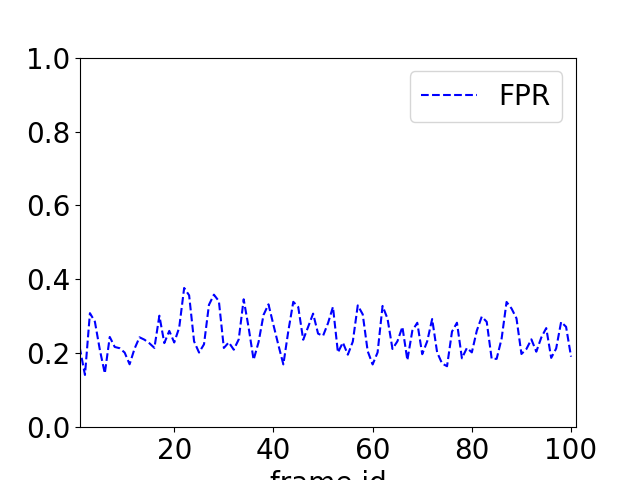}\\
          % \vspace{0.02cm}
          % \includegraphics[width=0.9in]{./TF2_6_s2.png}\\
          \vspace{0.02cm}
          %\caption{fig1}
      \end{minipage}%
  }%
  \caption{Comparison of $TPR$ and $FPR$ for Bs1 with different threshold. One point in the figure represents a frame. The x-axis is the frame id. The y-axis is the ratio value ranging from 0 to 1. The top line is $TPR$, and the bottomline is $FPR$.}
  \label{fig:bs1add}
\end{figure}

This experiment is carried on Gesture dataset recorded from DVS128. Since GF$_{2}$ method shows good performance on denoising the frame and the time consumption is also acceptable, we use it as a evaluating method for the other filters.
We calculate the $TPR$ and $FPR$ of the event-based filters. $TPR$ is the percentage of correct predictions for real events, and $FPR$ is the percentage of predicting a BA event as a real event. Fig.~\ref{fig:tfsnr} shows the results. The $FPR$ of all filters are low which is good to see, especially Bs2. These baseline filters rarely mistake the BA events defined by our criteria as real events, which suggests that our definition is approved by the baseline filters. The $TPR$ witness different distributions. Bs2 is the best. Bs3 shows the lowest $TPR$ which explains the reason for light outlines of objects as shown in figures in section~\ref{sec:visual effect}. We fund that Bs1 only pass about half percent of real events. We suppose the reason might be that the threshold for Bs1 is too low. So we adjust the threshold for Bs1 to 1ms. And the Fig.~\ref{fig:bs1add} shows the result. After increasing the threshold, it also shows good performance on $TPR$.

\subsection{time comparison}
\begin{figure}
  \centering
  \includegraphics[width=.8\linewidth]{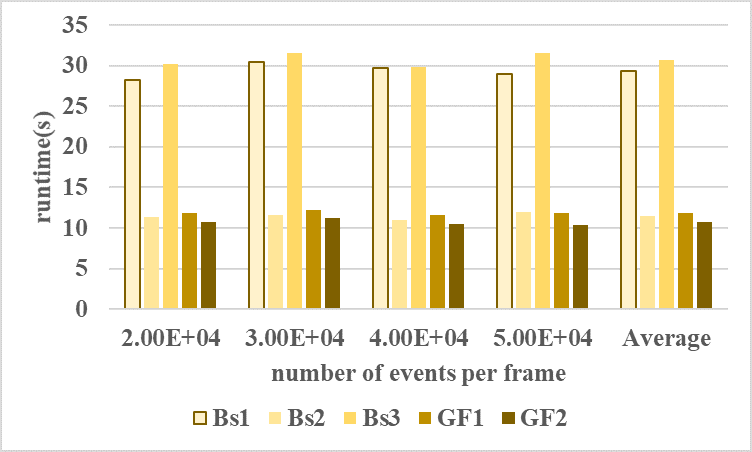}
  \caption{Runtime Comparison of different filters. We use 3 million events and different number of events to make a frame. The x-axis is the number of events per frame. The y-axis is the time consumption of filtering in total and the unit is second. }
  \label{fig:timefiltertime}
\end{figure}

Fig.~\ref{fig:timefiltertime} shows the time consumption of different filters.
We make several repeat experiments by using different number of events per frame with the fixed number of events (3 million) from a event stream. And with the same tendency, we can see that the Bs1 filter is 2.5x time consuming than GF$_{1}$ filter. This time reduction is achieved because the GF$_{1}$ only needs to write once and compute once. However, the Bs1 needs to write 9 times for updating the timestamps of 8 neighbors and the pixel itself, and compute once according to the process in \cite{Bs1filter}. We can see that GF$_{2}$ is similar to Bs2 in time cost. Also, GF$_{2}$ is similar to GF$_{1}$ in average time consumption because for GF$_{2}$, it also writes once and computes once for each coming event as well.

\subsection{Discussion}
One interesting behavior is demonstrated by the Bs3 filter. For Roshambo dataset, where each event has its own timestamp, Bs3 still works but it filters large amount of events as it shows the relative light outline of the object. However, for CeleX dataset, where up to a row of events share the same timestamp, it still works for the left part of the pixels but when the pixel is at the right side of the output, it has almost no filtering effect. This is especially clear in Fig.~\ref{fig:roshambo2}.

%the time difference between two real events in a pixel could be almost the same as the time difference between two BA events in a pixel. The performance of Bs1 and Bs2 support this too, even with neighbors in mind, the time difference of many real events cannot satisfy the fixed time threshold and the real events building up the outline of the hand disappear as the BA events.

\begin{figure}
\centering
\subfigure[fast-s2]{ % 6
    \begin{minipage}[t]{0.24\linewidth}
        \centering
        \includegraphics[width=1.2in]{./Bs2_12.png}\\
        \vspace{0.02cm}
        \includegraphics[width=1.2in]{./TF2_12.png}\\
        \vspace{0.02cm}
        %\caption{fig1}
    \end{minipage}%
}%
\subfigure[fast-s4]{ %32
    \begin{minipage}[t]{0.24\linewidth}
        \centering
        \includegraphics[width=1.2in]{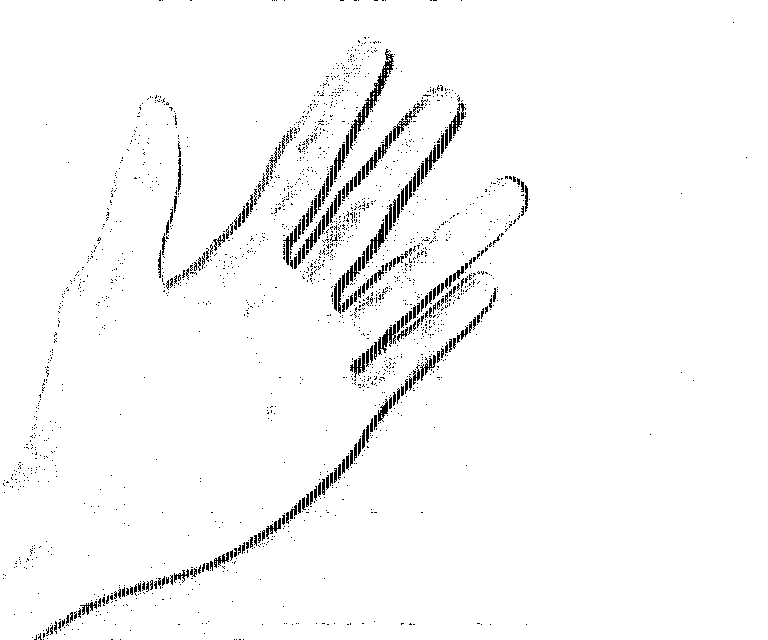}\\
        \vspace{0.02cm}
        \includegraphics[width=1.2in]{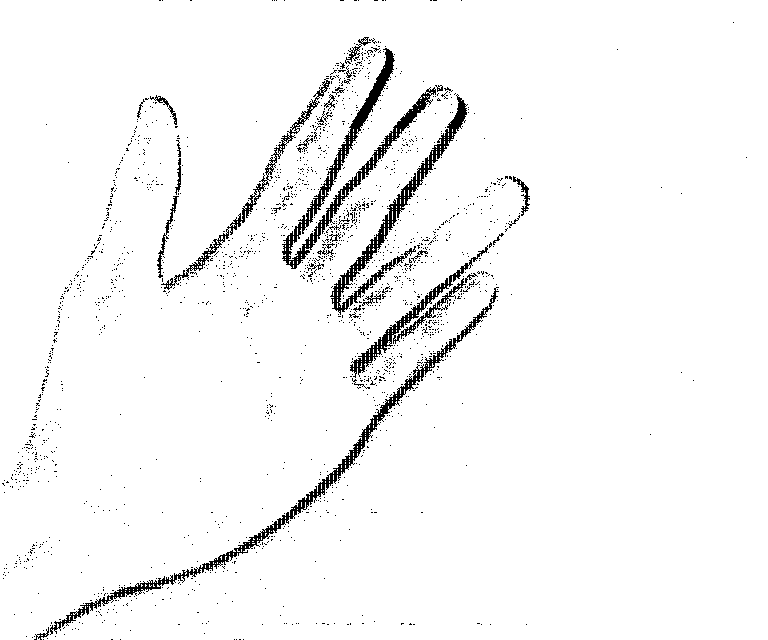}\\
        \vspace{0.02cm}
        %\caption{fig1}
    \end{minipage}%
}%
\subfigure[slow-s2]{ %75
    \begin{minipage}[t]{0.24\linewidth}
        \centering
        \includegraphics[width=1.2in]{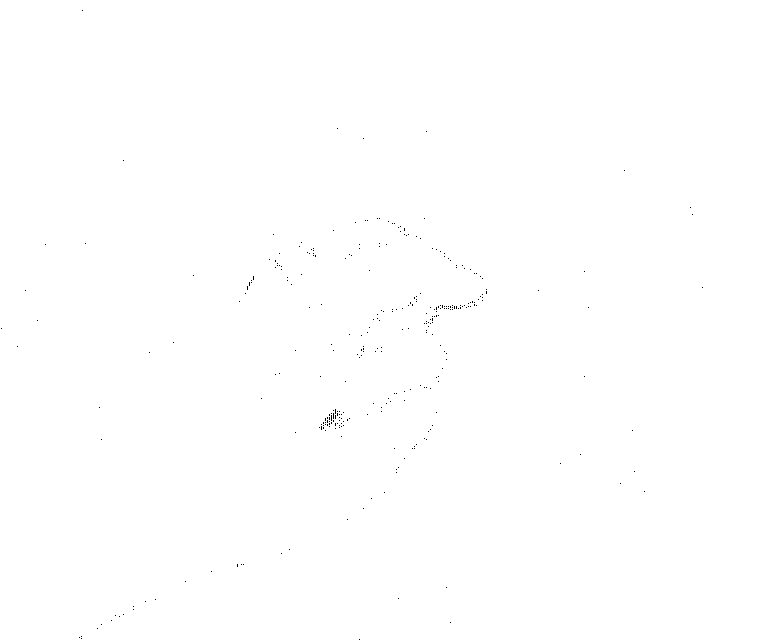}\\
        \vspace{0.02cm}
        \includegraphics[width=1.2in]{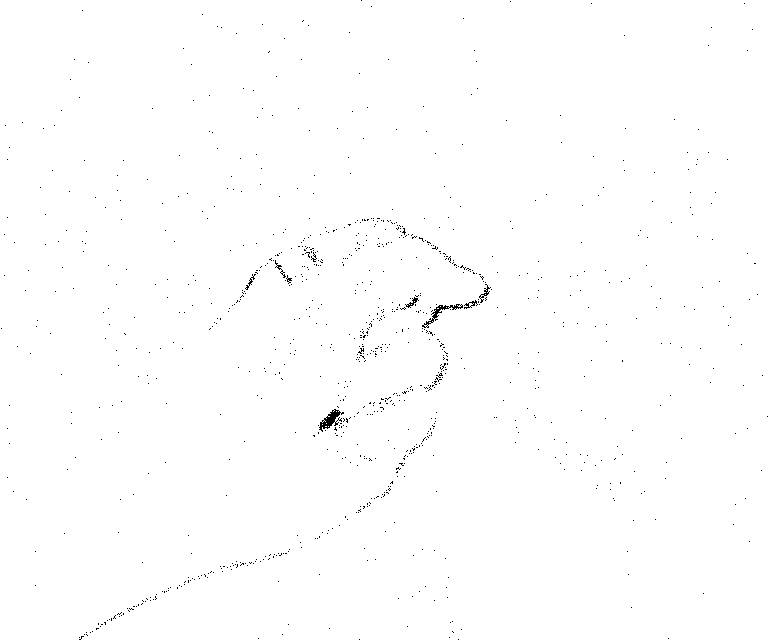}\\
        \vspace{0.02cm}
    \end{minipage}%
}%
\subfigure[slow-s4]{ %75
    \begin{minipage}[t]{0.24\linewidth}
        \centering
        \includegraphics[width=1.2in]{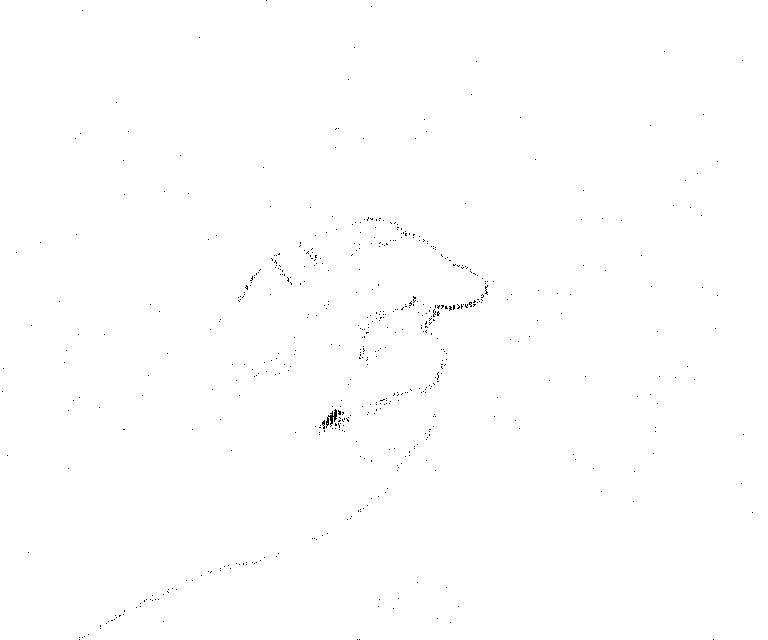}\\
        \vspace{0.02cm}
        \includegraphics[width=1.2in]{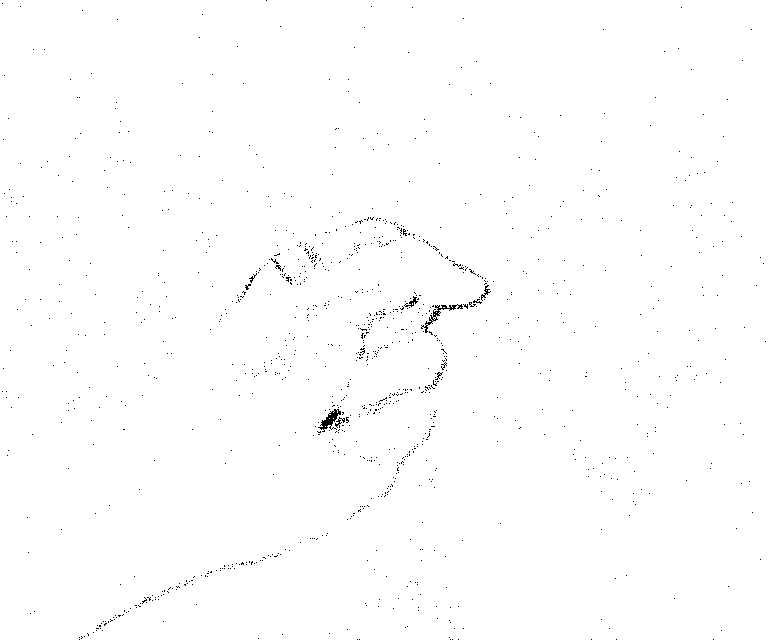}\\
        \vspace{0.02cm}
    \end{minipage}%
}%
\caption{Performance comparison between Bs2 and Time Filter with $s = 2$ and $s = 4$ on hand frames. $s2$ means $s = 2$. Fast means the hand is moving fast. Slow indicates the hand is moving slow. The subsampling window $s$ for Bs2 and Time filter are same. The top row is the Bs2. The bottom row is the Time Filter.}
\vspace{-0.2cm}
\label{fig:tfgroup1}
\end{figure}

We also make experiments on different subsampling windows as shown in Fig.~\ref{fig:tfgroup1}. We can see that the Time filter performs better than Bs2 with different windows, especially in slow movement scenarios.

% \subsection{Memory perspective}
% \subsubsection{Hardware}
%\vspace{-0.11in}

%\vspace{-0.32in}
\section{Summary and Conclusions}
%\vspace{-0.16in}
% Neuromorphic computing has attracted increasing attention among researchers working on artificial intelligence due to their energy efficiency and brain-like features. And
Neuromorphic event-based sensors have witnessed rapid development in the past few decades, especially dynamic vision sensors. These sensors allow for much faster sampling rates and a higher dynamic range which outperform frame-based imagers. However, they are sensitive to background activity events which cost unnecessary communication and computing resources. Moreover, improved noise filtering will enhance performance in many applications.
We propose a new criteria with little computation overhead for defining real events and BA events based on the space and time information of the event stream. We utilize the global information rather than the local information by Gaussian convolution.
The experimental results show that the proposed criteria shows good performance on denoising and run very fast.

% \begin{acks}
% \end{acks}

% \appendix
% \section{My Appendix}
% Appendix sections are coded under \verb+\appendix+.
%
% \verb+\printcredits+ command is used after appendix sections to list
% author credit taxonomy contribution roles tagged using \verb+\credit+
% in frontmatter.

% \bibliographystyle{cas-model2-names}
\bibliographystyle{unsrt}
% \bibliography{cas-refs}
% \bibliographystyle{ACM-Reference-Format}
\bibliography{filter-base}

% \end{spacing}
\end{document}